\pdfoutput=1
\documentclass[11pt]{article}
\usepackage{UF_FRED_paper_style}
\usepackage{verbatim}
\usepackage{indentfirst} 
\usepackage{subfloat}
\title{Account for Neuronal Representations from the Perspective of Neurons} 

\author{Chiyin Zheng\\
\\
        \small\text{Institute of Psychology, Chinese Academy of Science}\\
        \small\text{Beijing, China}\\
    \href{mailto:zhengcy@psych.ac.cn}{\texttt{zhengcy@psych.ac.cn}} 
    }
    
\date{\today}

\begin{document}
{\setstretch{.8}
\maketitle
\renewcommand{\baselinestretch}{1.2}\begin{abstract}
Mounting evidence in neuroscience suggests the possibility of neuronal representations that individual neurons serve as the substrates of different mental representations in a point-to-point way. Combined with associationism, it can potentially address a range of theoretical problems and provide a straightforward explanation for our cognition. However, this idea is merely a hypothesis with many questions unsolved. In this paper, I will bring up a new framework to defend the idea of neuronal representations. The strategy is from micro- to macro-level. Specifically, in the micro-level, I first propose that our brain’ prefers and preserves more active neurons. Yet as total chance of discharge, neurons must take strategies to fire more strongly and frequently. Then I describe how they take synaptic plasticity, inhibition, and synchronization as their strategies and demonstrate how the execution of these strategies during turn them into specialized neurons that selectively but strongly respond to familiar entities. In the macro-level, I further discuss how these specialized neurons underlie various cognitive functions and phenomena. Significantly, this paper, through defending neuronal representation, introduces a novel way to understand the relationship between brain and cognition.
\noindent
\textit{\textbf{Keywords: }%
neuronal representation; neural coding; localist coding; associative learning; hebbian learning; brain and mind.} \\ 
\noindent

\end{abstract}
}


\section{Introduction}
Researchers have long been curious about how concepts are represented inside our brain and it is speculated that individual neurons can serve as the substrates of mental representations in an explicit manner\cite{RN448,RN465,RN503,RN640}. This idea is to a certain degree similar to the so-called “grandmother cell” hypothesis\cite{RN460,RN461}, which may be the most controversial idea among neuroscience. Although this idea has been criticized for its obvious weakness\cite{RN501,RN833,RN860}, mounting empirical data from single neuron records are suggesting the possibility of its soft versions. Hubel and Wiesel’s famous experiment\cite{RN542} shows that neurons in the primary visual cortex are selectively responding to different orientations and ocular. Although Hubel and Wiesel themselves didn’t agree with the notion of grandmother cells\cite{RN461}, further researches following their methodology in vision study have found neurons selectively responding to more complex things in higher visual regions. Neurons in MT were found to be selectively responding to certain dimensions of motion and neurons in IT regions were found to be selectively responding to abstract features\cite{RN569,RN839,RN841,RN647}. More famous examples came from recordings in the hippocampus and MTL where neurons selectively responding to the concepts of celebrities like Jenifer Aniston and Marilyn Monroe were found\cite{RN469,RN636}. Besides vision studies, specialized neurons can be found across our brain. Researchers in various fields have found grid cells encoding space information in the posterior hippocampus\cite{RN546}, engram cells encoding specific events in the dentate gyrus\cite{RN651,RN258}, and mirror neurons encoding specific intention and action in parietal lobes and frontal lobes\cite{RN581,RN582}. Although these neurons differ in many aspects, they all similarly exhibit high selectivity and only respond to specific entities. More interestingly, their activities can be immediately related to the conscious contents at the moment\cite{RN830,RN1633}. All these facts suggest the possibility that neurons are specialized and serve as the substrates of mental representations in an explicit manner.\\
Neuronal representation shows its further power when combined with associative learning\cite{RN457,RN587,RN755,RN459,RN748}. If different individual neurons do represent different features or concepts, then by connecting specific neurons with synapses, one can readily create complex mental contents. This idea at least can be traced back to Donald Hebb. In his groundbreaking monograph, \emph{the Organization of Behavior}\cite{RN545}, Hebb brought up a putative and simple rule, which is known as “neurons fire together, wire together" today, to account for our ability of learning during our interaction with the environment. In his opinion, learning relies on the formation of the connections between neurons responding to different concepts, for example, corners of a triangle. As these entities frequently show up together, their corresponding neurons will fire together and hence wire together, forming a neural assembly for complex entities like a triangle or sequence for actions. The premise of this account is that different neurons do carry specific meanings in a context-independent manner so that they can be consistently activated by their targets. In turn, if this premise is accepted, its embrace with Hebbian rules, which has been empirically verified today, will provide a straightforward explanation for learning and memory. Kandel has demonstrated it on aplysia already\cite{RN554}. Similar results have later been obtained in memory studies on higher vertebrates. Researchers have found that memory can be rapidly created and represented as the connections between neurons representing its different components become strengthened\cite{RN755}. And thanks to the recent development of optogenetic methods, neuroscientists can directly create or alter memories by selectively activating neurons encoding different memory components\cite{RN810,RN525,RN365}, which doubtlessly provides more direct evidence. Besides learning and memory, a variety of other phenomena, including conditional reflex, priming effect\cite{RN549}, and false memory\cite{RN553} can be explained by the combination of neuronal representations and associative learning. \\ 
Taken together, these results suggest that individual neurons can represent different concepts and their connections underlie complicated thoughts\cite{RN361,RN748}. These straightforward ideas, if proved correct, will much excite researchers, especially those believing in materialism, as they can find the concrete structure for many abstract phrases. Questions plaguing psychologists and philosophers for centuries can be easily answered. Besides theoretical significance, these ideas also find an interface between the mental world and the physical world, which can promisingly promote or reshape many industries including education and psychiatry.\\
However, the problem is, are these straightforward ideas truly reliable? Are these neurons selectively responding to Aniston or Monroe really their representations? Are their connections really the substrate for complicated thoughts? While indicated by mounting empirical results and has actually been widely used to explain various phenomena, they are still no more than hypotheses. Moreover, while many researchers believe that neurons are specialized, many others hold different views about coding and connections\cite{RN539,RN556,RN538,RN694,RN543}. Particularly, a list of disadvantages of coding schemes based on that individual neurons encode concepts in a point-to-point manner, including the combinational explosion and vulnerability, have already been pointed out long ago. These convincing criticisms make it hard to believe that our complicated mental world is really built upon these simple and straightforward ideas even with mounting evidence in support of them.\\
But how to prove that neurons are really specialized and selective? The data from single-neuron recording, though provides clues, is insufficient. First, the data is far from enough since it is impossible to test every neuron with every stimulus not even considering the dynamic property of our brain\cite{RN846}. No one can assure these cells are truly selective and dedicated based on this limited data. Second, while there is evidence for neuronal representations, there are more adverse observations where neurons are found to be responsive to a range of stimuli. Lastly, although neurons with high selectivity for stimulus is predicted by these ideas, it is not logical to prove them with the finding of these neurons, or it will fall into circular reasoning, especially when the data can be explained in other ways\cite{RN543}. Hence, empirical data must be accompanied by analyses from other perspectives. Cognitive phenomena that can be well explained by these ideas surely promote supporters’ confidence, yet they cannot be used as evidence because it is also clearly circular reasoning. Computational models based on the idea of competitive learning seem to be able to explain how neurons become selectively responding to certain features\cite{RN564,RN498,RN562,RN491,RN664}. And units in convolution neural network, which is inspired by the human vision system, are reported to be able to extract specific features like shape and texture from training data. Yet they can only offer insights but cannot serve as evidence. We cannot prove birds fly on wings by finding planes fly on wings. Finding feature-selective units in computational models cannot directly prove neurons in our brains are also selective. \\
Nevertheless, besides accounting for the inconsistency among empirical data and answer a range of theoretical concerns against neuronal representations, two categories of questions as Marr pointed out\cite{RN1375,RN464} are critical. The first category is the why questions. This category of questions includes why these neurons are really representations for concepts, why we should represent the world in this way, and why it is fair to call the neuron responding strongly and dedicatedly to roses 'rose neuron'. The second is the how questions. This category of questions includes how these neurons emerge during learning, how can they represent infinite objects, and how they further contribute to our cognition. \\
In this article, I will bring up a theory from a different perspective to support the ideas of neuronal representation by answering these two categories of questions. The strategy here will be Schelling’s way\cite{RN844} where I will first discuss neurons’ behaviors at the micro-level and then, discuss how neuronal representations at the macro-level can emerge from their behaviors. For the sake of simplicity, while the neuronal representation may be actually carried by a set of neurons or cortical columns, I will only use the word neuron for both a single neuron or a group of neurons. Besides, the usage of words like ‘entity’ or ‘concept’ will be quite casual here. They are not confined to their lexical meanings but used freely to refer to ‘things’. The reason for this casual treatment will be explained later.

\section{The basic assumption}
A basic assumption for the later discussion is that our brain favors and preserve more active neurons. Such a preference can be expressed by allocating more energy or other supports crucial for survival to neurons firing more frequently and will push neurons to pursue more excitation to fire frequently and strongly. Several points must be noted. First, this can be interpreted in a way similar to the Evolution Theory that only the neurons fire frequently can survive. Neurons don't have to bear an intention to fire more in their soma just like giraffes don't have to be intentional to have a long neck. Second, living cells as neurons are, it is their instinct to take advantage for survival. Hence though not necessary, it is still plausible that there are mechanisms supporting neurons to initially pursue more excitation. Third, this is more of a general tendency instead of a strict law. While it suggests that neurons that fire more frequently can have a better chance to survive, it doesn’t deny the existence of neurons rarely or never fire\cite{RN810,RN811}. Although this rule has not been explicitly put forward ever before, its plausibility can be supported by the following arguments. \\
It can be first supported by facts from neuroscience and developmental neuroscience. First, it is observed that there is a tight coupling between neuronal activities and local energy metabolism, indicating that neurons’ activation can earn them energy supplies back. Second, the production of neurotrophins, which is essential for neurons’ survival and growth, is positively regulated by neuronal activity\cite{RN490,RN494}. Third, glial cells favor those active neurons and synapses, too. Specifically, astrocytes will enhance their interaction with neurons and synapses more active\cite{RN531,RN533} and mediate the elimination of those less active\cite{RN606}, microglia cells will engulf those inactive synapses and initiate programmed death in those neurons that fail to collect enough neurotrophin\cite{RN488,RN604}, and myelination, which involves oligodendrocytes, is also activity-dependent\cite{RN524,RN518,RN517}. And there are more other mechanisms that help to select active structures\cite{RN603,RN608,RN607}. Taken these facts altogether, it wouldn’t be baseless to conclude that our brain favors those more active neural elements.\\
Besides and beyond physiological evidence, such a preference can be plausible also because it is advantageous and necessary for our brain. For one thing, neurons that never fire can hardly exert any influence or contributes to our cognition but still consume energy and occupy space in our brain. Were there not mechanisms to eliminate these redundant neurons, our brain will be inefficient. For another thing, neurons firing more frequently demand more energy and support as generating spikes consumes much. Since creating new vessels and connections between glial cells and neurons take both time and resources in the actual world, it is reasonable for our brain to allocate more supports to neurons that are more active in advance.\\
Excitation is related to neurons' income. Neurons that gain more support for the brain by firing more frequently and strongly will have better chance for survival while the poor neurons will be eliminated. Taking the elimination of neuronal elements into consideration is not novel. Edelman has already brought it up in his Neural Darwinism Theory\cite{RN727,RN162,RN565,RN356}, where he postulated that the brain will create an excess number of neuronal elements and later preserve those valuable. Although the initial excess and the later removal are clear and supported by abundant empirical results, the definition and mechanisms of value are rather vague. The problem is while it is the neural elements that are to be selected, the value is in terms of the organism as a whole. This deviates from Darwin’s idea and is like saying that the giraffe is selected because its existence accords with the value of veld. Here, instead of throwing the question to subcortical neurons, the value for neurons is specified as their excitability. But on the other hand, since generating action potential consumes much energy and the energy supply for the brain is limited\cite{RN1643}, although neurons dream to fire more, there can only be a portion of neurons can fulfill their goal. The activity of neurons exhibits skewed distribution, Most neurons fire slowly while only a minority of neurons can fire frequently. That's to say, neurons, either initially or passively, have to compete for the limited chance of discharge and survival.
\section{The strategies}
Having assumed that the selectional pressure is upon neurons' excitability, it should be discussed how should a neuron react in response. What can be a neuron’s strategy to fire more? There are numerous observations in the brain whose functional roles for our cognition have been much asked and speculated. However, no one has ever what these mechanisms meant for neurons. Here, I will propose that many observations in our brain can function as neurons' strategies for survival. It should be noted that this sector’s discussion will have nothing to do with cognition or information processing, but is only about how these mechanisms help neurons to survive in our brain.

\subsection{Synaptic plasticity}
Synaptic plasticity has been very familiar to neuroscientists and psychologists since 1973\cite{RN573}. Generally, it is classified into 3 types, LTP, LTD, and STDP\cite{RN433}. Although the exact implementations can vary across the brain, it can still be generally concluded that LTP means synapses receiving high-frequency inputs and resulting in high amplitude excitatory postsynaptic potential(EPSP) will be facilitated while LTD is the inverse process. While their molecular details have been much understood\cite{RN576,RN575}, it’s still an interesting question that what's the logic behind these complicated processes. Especially, it’s tempting to ask why the frequency of inputs and the intensity of EPSP will matter. Why a neuron should strength its connections with neurons firing strongly and causing strong depolarization but not the reverse? Here, I propose that this is neurons’ investment strategy for more excitation and therefore, a greater chance of survival.\\
Since neurons’ goal is to fire more frequently, it is not hard to understand why the intensity of EPSP will be decisive. Another question is how should neurons adjust their connections so that they can fire earn more excitation. The connections of a neuron are like its investments to its upstream neurons. It is reasonable to suppose that the excitation a neuron gains from one upstream neuron is positively correlated both with the firing intensity of this upstream neuron and its synaptic strength. For simplicity, the synaptic revenue can be assumed as the product of synaptic strength and the pre-synaptic activity and hence the total revenue a neuron gains from its upstream neurons can be written as the dot product of the input vector and its weight vector. But meanwhile, maintaining a strong connection in the real biological world inevitably costs much. To gain more excitation from its upstream neurons, a neuron has to invest properly. And it's obviously a reasonable strategy to allocate investment in proportion to the firing rate of upstream neurons. More connection investment should be added to neurons firing strongly and frequently to maximize its profit from these blue chips. On the contrary, if a strong connection has been established yet the input is far below the expectation, then it should withdraw its investment to it. Based on such a strategy, if an input pattern, which means several upstream neurons are consistently firing together, presents frequently, the neuron will accordingly adjust its weight vectors to be in alignment with this input vector. This is firstly a safe asset allocation strategy for neurons. Then for one thing, if the norm of the weight vector is fixed, then such an alignment can maximize its revenue from this pattern. And for another thing, the more frequently this pattern appears, the more total revenue a neuron can gain in a long term. \\
Although the exact implementation of this strategy can be much more complex, the essence of synaptic plasticity is always about how can neurons allocate connections appropriately to gain more excitation from their upstream neurons and become more active. Following this basic logic, it wouldn't be hard to answer why neurons should strengthen connections with synapses sending high-frequency inputs and why there isn’t a fixed reversal point between the occurrence of LTP and LTD but it is highly dependent on the synaptic history. \\

\subsection{Inhibition}
A large number of neurons in our brain are inhibitory neurons\cite{RN693}. Yet again, while their physiological properties have been well studied\cite{RN534,RN364,RN406}, its functional role remains a mystery. Since these neurons receive excitatory inputs from excitatory neurons but in turn inhibit the activation of excitatory neurons, their job is naturally thought to be maintaining the cortical balance and prevent excitatory neurons from overheating\cite{RN364}, but this is merely speculation. Here, I propose that the functional role of inhibitory neurons is not to hinder excitatory neurons, but is to help their seeming counterpart. Their existence is the second strategy on the neurons’ survival guide.\\
The implementation of the first strategy means a neuron will be attracted by a frequent input pattern and adjust its connections to be able to fire to it. Yet the thing is, this beneficial pattern will attract more than one neuron. When this pattern appears, multiple neurons will fire concurrently and each will take a share of the benefit. Hence, for a neuron trying to gain even more benefit, besides adjusting its synaptic weights precisely, another strategy is to deactivate other neurons, especially those firing together with it, to prevent them from sharing the benefit. To achieve so, an excitatory neuron will connect to inhibitory neurons and when this neuron is activated, its activation will be spread to the inhibitory neurons it connects with\cite{RN667}. In turn, these activated or recruited inhibitory neurons will suppress a population of excitatory neurons for it\cite{RN622,RN621,RN616}. Each neuron participating in the competition will recruit inhibitory neurons, which will progressively increase local inhibition tone and deactivate neurons less excited. Finally, only neurons firing strongly enough can survive the competition and take the benefit\cite{RN621,RN619,RN406}. Therefore, although an excitatory neuron will also be deactivated by the inhibitory neurons it recruited, it can maximize its benefit by preventing its peers from discharge if it fires strongly enough.\\
Contradictory to the intuitive thought that inhibitory neurons are competing with excitatory neurons, they are actually the cooperative partner, or more analogically, the arms of excitatory neurons and help the latter to compete with each other. This relationship can explain why the maturation of inhibitory neurons will follow that of excitatory neurons and is regulated by BDNF from their connected excitatory neurons\cite{RN242}. This new relationship can further help to understand a range of observations. Since this cooperation originates from excitatory neurons’ desire to monopolize a certain input pattern, it is not hard to understand why the strongest lateral inhibition will occur among neurons that fire together\cite{RN620}, why the inhibition will dynamically increase with excitation\cite{RN639,RN611}, and why the distribution of inhibition is found to be precisely matched with that of excitation\cite{RN712,RN159,RN492}. All these facts follow the logic that inhibitory neurons are recruited for the competition among excitatory neurons.\\

\subsection{Synchronization}
The strong competition in turn forces a neuron to take the third strategy to integrate more excitation within a narrow temporal window so that it can break through the strong inhibition. The implementations of this strategy can be different according to the relationship between neurons, but its essence is always simultaneity.\\
For a neuron at the downstream, now that the inhibition is considered, the adjustment of synaptic strength to its upstream peers can't solely accord to their firing rate but also the timing. To cope with the strong competition, the spikes from upstream neurons must arrive within a narrow time window to ensure the integrated excitation is strong enough to overcome the suppression\cite{RN813}. The connections to neurons whose inputs don’t arrive together with the main force shall be weakened. And the connection budget will be concentrated on upstream neurons whose spikes arrive within the narrow time window to maximize its excitation\cite{RN1639}. Consequently, a neuron executing this strategy will be highly selective to input patterns and fire strongly when very specific patterns appear\cite{RN589,RN601,RN649}.\\
To receive more concurrent inputs from upstream neurons to break through the inhibition, it has to inform its upstream peers of its firing and more importantly, attract them to make connections with it. To achieve so, it has to establish backward connections with upstream neurons firing concurrently and send excitation backward both for synchronization and invitation. As for the upstream neurons, they will be glad to accept the invitation and connect with downstream neurons firing strongly and simultaneously with them because integrating both backward and forward inputs is beneficial for their survival as well\cite{RN449,RN447,RN450}. In this way, reciprocal or reentrant connections will be established between neurons at different levels firing concurrently\cite{RN697,RN745,lamme2000distinct,RN569}. The bidirectional connections between these neurons further help to synchronize their activities, which in turn allows these neurons to integrate explosive inputs within a narrow temporal window and have advantage in the local and global competition. Hence, it can be seen that the word ‘reciprocal’ is quite appropriate as the mutual support provided by bidirectional connections can benefit neurons at both levels.\\
Cooperation can also happen between neurons in the same layer. Compared with reciprocal connections between layers, lateral connections are much weaker in our brain\cite{RN633,RN632}, but the principle is the same. If two neurons can be activated by patterns that frequently show up together, then the lateral excitatory connection between them can benefit each other\cite{RN526}. Otherwise, lateral connections will be useless if the lateral reinforcement cannot be integrated with the upstream inputs. The stronger connections between neurons activated together also means they will be more prone to fire together spontaneously\cite{RN1651} and excitation can flow between them. What’s interesting here is that the competition is also the strongest between neurons fire together as discussed, which indicates that there can be highly dynamic and complex relationships between them\cite{RN628}. Neurons firing concurrently will huddle together when the situation becomes hard by synchronizing their firing for mutual support, but will also break up and compete with each other to for a bigger share of the benefit when abundant inputs come. Such a Schopenhauerian relationship between neurons results in the dynamic size of neural assemblies and neuron's firing property\cite{RN1655},\\
All these implementations comply with the Hebbian rule\cite{RN635}. But unlike Hebb brought up his rule abruptly as an assumption, here, it is called 'strategy' for helping neurons win the competition for survival. If the inhibition is weak, then the timing doesn't have to be that strict\cite{RN634}. Otherwise, when the competition becomes fierce, only neurons that find and connect with reliable allies can survive. The reliability is reflected by the punctuality because the strong inhibition requires neurons to gather more excitation within a narrow window, which relies on the punctuality of reinforcements from directions. This explains why timing is so decisive for making connections.

\section{How neurons become representations?}
In short, neurons' strategies for survival are investing properly, repelling competitors and finding reliable allies. And it's time to show how can neurons become neuronal representations after carrying these strategies. To make the discussion easier, I will only demonstrate it on two layers of neurons. And before it starts, the lower lever is stabilized, meaning that a certain stimuli, for example, rose, will elicit a relatively consistent activation pattern on this level across trials. by contrast, the upper level is still chaotic and plastic, which means 1) the activation pattern of this layer elicited by rose can vary from trial to trial, 2) the connection weights of neurons in this layer are randomly initialized and weak, 3) both excitatory and inhibitory neurons are immature.\\
As brought up ahead, the firing pattern of the lower level can be seen as an n-dimensional vector where the value of each dimension is the firing rate of each neuron. These lower-level neurons can in this way create an n-dimensional \emph{conceptual space} where each point in it represents a firing pattern. Although the space is almost immense\cite{RN768}, the number of stimuli one encounters frequently in one’s short life is limited. These frequently present stimuli become some scattered points in this space and serve as the \emph{center of gravity}\cite{RN491}. And for a neuron in the upper level, input patterns bringing successful discharges constitute its \emph{excitation territory}. This term excitation territory corresponds to the well-known term tuning curve but is viewed from different directions. For a person, if roses frequently appear in his or her life, then rose is a center of gravity. And if a neuron in the upper level can be activated by a rose, then the rose is within its excitation territory(figure 1). \\

\begin{figure}[H]
    \centering
        \includegraphics[scale=.85]{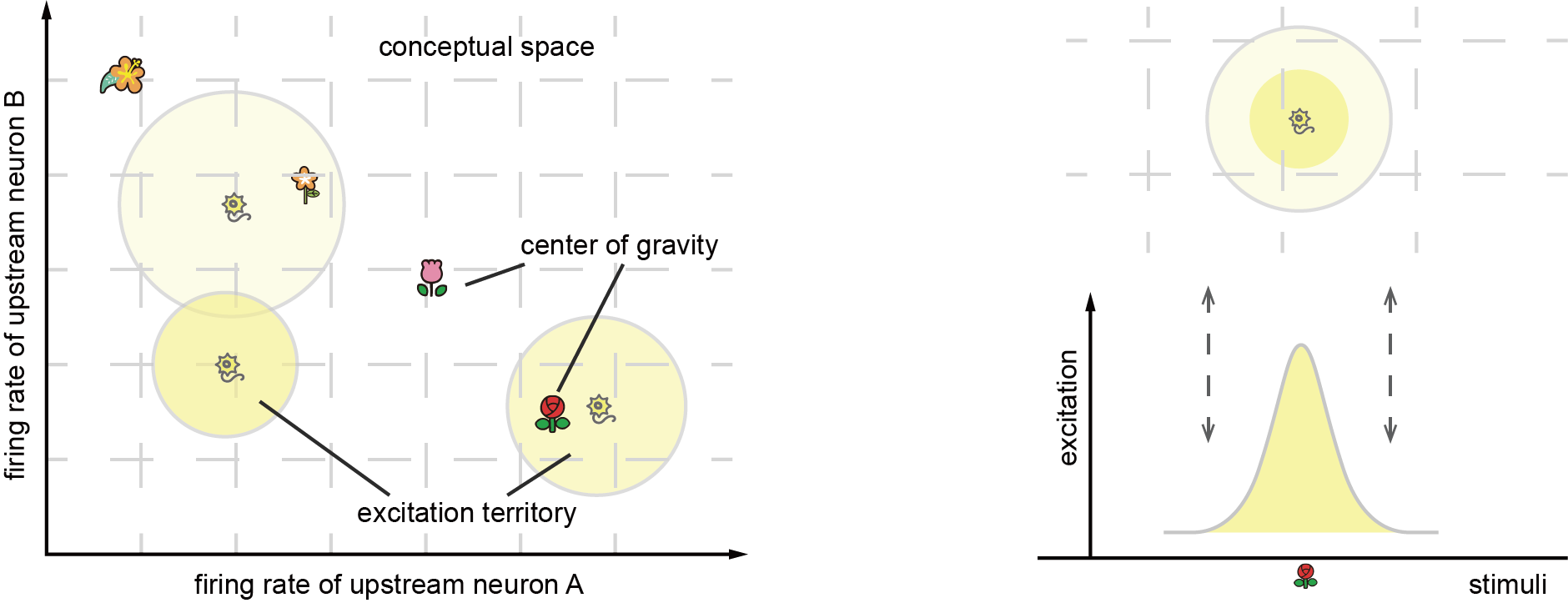}
    \caption{Left: the conceptual space is spanned by the firing rates of upstream neurons. Two dimensions for demonstration but will be n dimensions where n is the number of upstream neurons. Each point in this space is a firing pattern. Frequent patterns like flowers serve as the centers of gravity. A neuron's excitation territory represents its response to different inputs. Neuron's connections with upstream neurons dictate its response to inputs and hence its territory. Right: the excitation territory corresponds to the tuning curve.}
    \label{fig:1}
\end{figure}
The excitation territory of a neuron is apparently defined by its connections with upstream neurons as connections dictate how strong this neuron can respond to certain input patterns. In particular, the weight vector of a neuron, which is also an n-dimensional vector, marks the center of its territory. A neuron's adjustment of connections means both moving in the conceptual space and claiming new excitation territory. What a neuron is going to do is to obtain and exploit a piece of excitation territory in the conceptual space so that it can gain enough chance of discharge and survive upon it.\\

\subsection{First stage: emerging}
Since the inhibition is low in this first stage, the dominating strategy will be the first strategy. Two things hence will happen. First, threshold caused by the inhibitory neurons is still low, excitatory neurons can fire readily to inputs and will establish massive connections with upstream neurons\cite{RN405}. Second, to earn more excitability, it is beneficial for a neuron to align its weight vector to the firing pattern of a center of gravity. Consequently, neurons in this stage will be attracted by centers of gravity and possess large excitatory territories around them. Its large territory means it will respond to a range of similar stimuli. Meanwhile, since a number of neurons are attracted by a same center, their excitatory territories will largely overlap and all these attracted neurons can be activated by this stimulus. As multiple neurons will response to multiple stimuli, the region as a whole will exhibit strong responses to stimuli that just frequently appear. Such low selectivity and high excitation of both individual neurons and the region as a whole are in agreement with observations of the initial stage of learning\cite{RN843}.\\
To demonstrate, if flowers, including rose, lotus, tulip, and others, are frequently presented to someone, these stimuli will all serve as centers of gravity in the conceptual space created by feature neurons at lower levels. Multiple neurons at the upper level will be attracted to the activation patterns elicited by these flowers and move by connecting with these feature neurons. These attracted neurons will now possess large territories and share overlapped excitatory territories around these centers. Since these flowers share many common features and are quite similar, their induced centers will be adjacent in the conceptual space. In this case, several centers will fall within a same neuron’s territory and this neuron can be activated by several varieties of followers. In turn, the presence of one flower will activate a large portion of these attracted neurons.\\
This is the start of neurons' excitation rush. Numerous neurons will be attracted by frequently present patterns. Among them, neurons that have successfully executed the first strategy by allocating their connections properly according to the frequent input patterns can earn more excitation and hence more supports from the brain than their peers, which allows them to establish more and stronger connections and grow larger\cite{RN490,RN624}. This becomes positive feedback as more connection budget in turn allows these neurons to fire more strongly and gain even more excitation. These neurons will hence soon stand out from their peers. More importantly, they now have excessive excitation and thus can take the second strategy by recruiting inhibitory neurons to further maximize their benefit.
\subsection{Second stage: competition}
While the first stage can be described as a wild stage, the second stage is a cruel stage. To repel other neurons from their excitation territories and take the benefits alone, emerging neurons start to connect to inhibitory neurons and support their growth, which triggers the fierce competition among neurons once sharing overlapped excitation territories. The neuron that can fire the strongest will win the control of the disputed territory and continue to enjoy the excitation from it, while the rest neurons cannot benefit from this space as they are submerged by the inhibition.\\
To be the strongest neuron to win the competition, the third strategy shall be taken. Firstly, a neuron has established massive connections in the first stage and hence possesses large excitation territories. However, most of these connections are useless now because the conceptual space to which these connections correspond has been occupied by neurons that are more strongly activated. In reaction, this neuron will cancel these redundant connections\cite{RN721,RN734,RN733} and stop wasting budget on these occupied spaces. Then, it should concentrate its budget on upstream neurons firing together, which means these upstream neurons are highly probably activated by a same object. By wiring intensively with these neurons and integrating inputs arriving within a narrow temporal window, this neuron can fire explosively and beat other competitors. To cope with the even stronger inhibition, it has to select a specific firing pattern of this group of upstream neurons and rather precisely adjust its connections according to  it\cite{RN758,RN768} to maximize its excitation when this pattern appears. Finally, it has to cooperate with neurons that fire concurrently with it in other directions by establishing reciprocal connections with them so that it can receive timely reinforcements from them to ensure its advantage.\\
With all these measures taken, a neuron, having discarded most of its territory claimed in the wild first stage, can promisingly occupy a condensed territory in the conceptual space, which corresponds to a narrowed tuning curve. A neuron will keep silent in most situations but fire non-linearly strong to specific inputs\cite{RN601,RN625,RN405,RN626}. Meanwhile, since it has to wire reciprocally with peers at the same layer, it may also respond to some related inputs\cite{RN756}. But this doesn't undermine its selectivity. With each neuron controlling a condensed territory, the conceptual space is now divided up precisely and the overlap among neurons is now minimized. A specific input can only activate a small number of neurons while others are submerged by the strong inhibition, which mirrors the so-called sparse coding\cite{RN845,RN696}.\\
In this stage, neurons once can be moderately activated by tulip can no longer fire to it because it's now inhibited indirectly by the neuron more strongly responding to tulips. To survive, it has to make an all-in to upstream neurons activated concurrently and consistently by roses, so that it can integrate explosive inputs arriving within a narrow window to break through the strong inhibition. In this way, it becomes a neuron that selectively and also strongly responds to roses but keeps silent to other flowers. Similarly, the rights for firing to carnation or lotus will be occupied by other neurons that adjust their connections properly. Furthermore, since a rose may usually accompany certain people or places, this rose neuron will also connect with these neurons for these people or places. Their activation will be spread to the rose neuron so this rose neuron can moderately respond to these people and places. But it is still a dedicated rose neuron. 
\subsection{Third stage: stabilization}
The second stage is a long-lasting and fluctuating stage. Although a neuron successfully executing three strategies will have an advantage in the long term, its lead is not instantly established. There are be many contenders with comparative excitation in early stage competing with this neuron for the territory. It takes time for it to gradually repel these contenders. During this gradual process, its territory is not stable. When it continues to fire, its excitation will attract more upstream neurons to establish stronger connections with it and hence it can expand its territory. However, claiming disputed territories then put it at a disadvantage. And when the input decreases like during the sleep, it has to retreat by canceling improper connections\cite{RN894,RN895}. Nevertheless, after thousands of times of establishing new connections and canceling improper ones, there will finally be neurons that have found the most proper connections and hence tightly controlled their excitation territories. And the cruel stage will then converge to a harmonious stage.\\
In the last stage, with multiple neurons having comparative excitation, the clashes will frequently occur since each of them has the chance of winning. The competition takes a long time and the result can be highly uncertain\cite{RN406}. Moreover, recruiting inhibitory neurons inevitably consumes much energy. But in this stage, the neuron that finally survives has wired most properly and will fire dominantly strong to its target, while other neurons won’t even attempt to contest with it as they know it’s totally a waste now that the gap of excitation is that large now. In this way, the consuming process of competition is avoided. A neuron thus can tightly control its territory as no one can compete with it in its zone. But it also means it cannot expand its territory because its neighbors have also tightly controlled their territories and the stable borders between these survivors will have to be settled. Being stable is not equal to being static, a neuron will still expand its territory after being excited, but it will soon be inhibited and pushed back by its neighbors\cite{RN638,RN698}. Hence, a neuron will be permanently confined to its zone even though its territory is still highly dynamic.\\ 
It is in this stage that the remaining neuron selectively responding to roses becomes a rose neuron. There have been be more redundant rose neurons in early stages, but most of them lose their qualification as the competition becomes increasingly tough\cite{RN1653}. Nevertheless, the neuron that survives the cruel knock-out stage will keep responding strongly, selectively, and dedicatedly to roses, and hence becomes the so-called gnostic neuron in effect. As the direct competition is much avoided in this stage, the efficiency of the region is much improved. Less time and energy will be consumed\cite{RN721,RN734}. But as the borders are set down, the plasticity is largely reduced and new learning becomes hard\cite{RN690,RN360,RN615,RN689}. These concurrent changes are in agreement with a range of developmental and learning facts(figure 2)\cite{RN405,RN400,RN699,RN407}.\\
\begin{figure}[H]
    \centering
        \includegraphics[scale=.85]{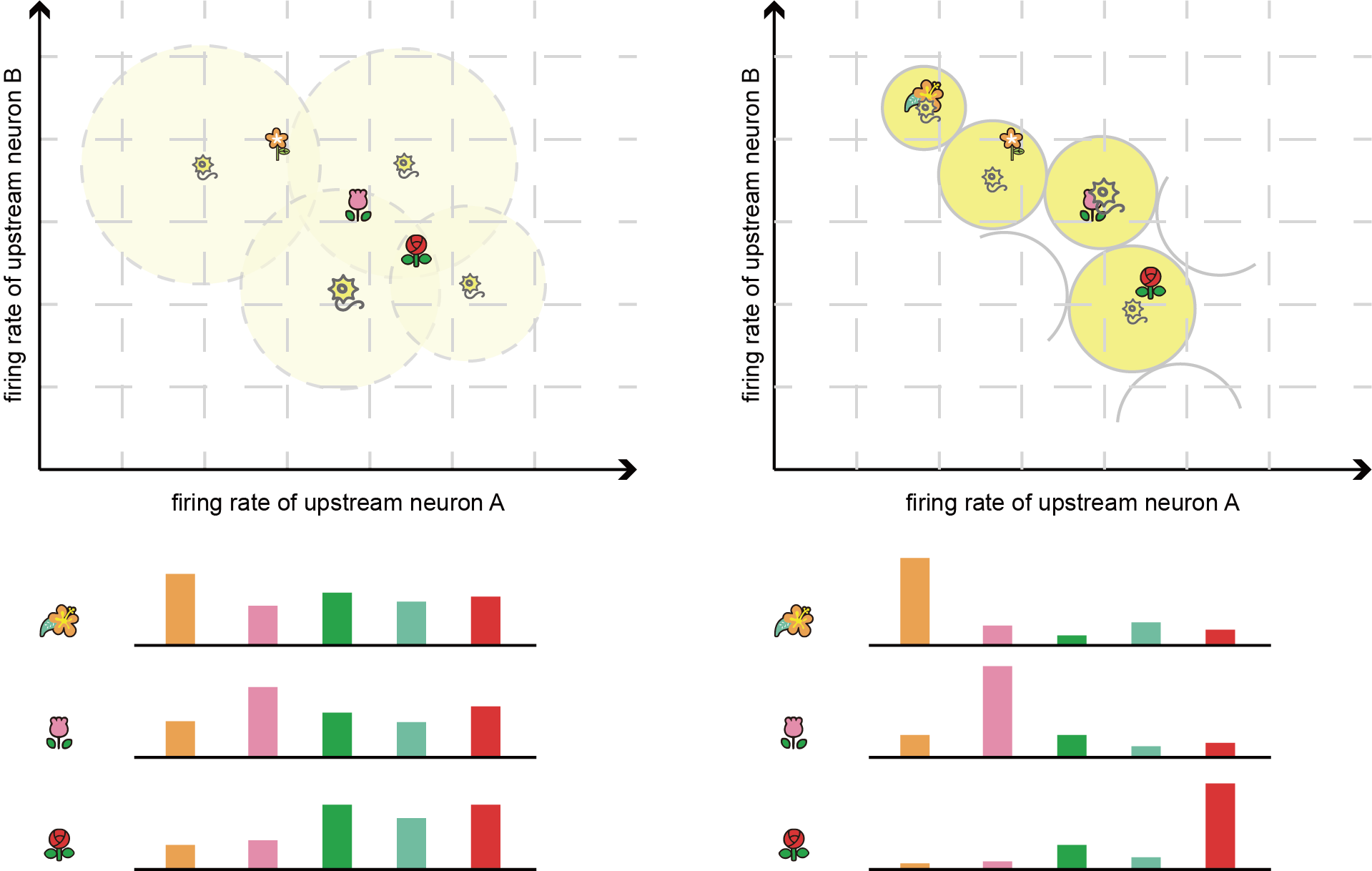}
    \caption{Left: at the early stage, each neuron covers large territory and their territories are largely overlapping. Each flower can activate many neurons while each neuron responds to many flowers. The region is very excitable but less selective. Right: after the stabilization, each neuron controls condensed territory. Each flower can activate only one of these neurons and each neuron fires selectively but also strongly to one flower. The activity of the region becomes sparse but informative.}
    \label{fig:2}
\end{figure}
Importantly, as each neuron has become a dedicated detector of its targeted input and interferes with others as little as possible after stabilization, the same stimulus will elicit consistent patterns across trials in this layer. This consistency echoes with the precondition put forward at the start of the section and will be the precondition for the development of the next level. Therefore, although the discussion here only involves two layers, it should be noted that this narrative can be recursive. This is also in agreement with the developmental sequence of our brain where different regions of the brain mature in accordance with a functional hierarchy and primary regions will mature before higher-order ones\cite{RN728}. However, it should also be noted that the narrative here is only for neurons attracted and participating in the competition. There can also be neurons that are not attracted by centers of gravity and live on and move around on the vast but barren conceptual space where centers of gravity are sparse\cite{dragoi2003}. They will fire weakly to a number of inputs and retain to be highly plastic. Analogically, these neurons can be named as \emph{nomadic neurons} while neurons discussed above can be named as \emph{settled neurons}. That’s also to say, not all neurons, but only neurons for frequently present entities will exhibit high selectivity. 

\section{Why specialized neurons?}
The answer for why neurons will be specialized is largely embedded in the demonstration of how neurons become so during the interaction with their synaptic environments. However, there have been many theoretical concerns and therefore objections to the idea that concepts can be represented by individual neurons. These doubts all sound reasonable. Therefore, efforts have to be taken to answer these doubts, which is a crucial part of the proof for the neuronal representations.

\subsection{Innate reasons}
The inevitability of neurons' specialization roots in the plain fact that neurons are selfish as living cells. The usage of the word 'selfish', is partly like that in the famous book by Dawkins\cite{RN700}. A neuron is called selfish simply means all a neuron cares for is survival and will take all possible actions that are advantageous for its survival. It is also tempting to call neurons 'rational cells' for the same reason. It has been discussed above that how can executing their strategy for survival turn them into Gnostic neurons, yet some issues still bear more discussion. And it should be discussed what can this simple notion that neurons are alive tell.\\
The neuron in our brain is blind to the outside world and all they can 'see' is the neuron transmitters they receive. It can't see the whole picture of rose, let alone appreciating its aesthetic value and understanding its cultural connotation. It seems impossible for neurons to become the inner representations of outside entities now that they completely don’t know what exists outside\cite{RN546,RN871,RN887}. This is correct of course. Yet paradoxically, the reason why a neuron can become a rose neuron despite its ignorance of the outside world and lack of grounding is that it doesn’t care for being a rose neuron. Their goal is simply firing frequently and strongly. A neuron is only responding to its upstream neurons instead of any objects in the outside world, yet the recursion of our brain allows it that the presence of roses will activate a group of upstream neurons consistently. This neuron's desire for discharge and a place in the conceptual space will direct it to wire with these neurons concurrently and frequently activated and take advantage of this coincidence to fire. It doesn't need a teacher to judge whether it makes a correct representation because its goal is not representing roses, but the discharge per se. Its pursuit of excitation will automatically drive it to wire more 'correctly' to respond strongly to the pattern elicited by roses. In this way, although it cannot see roses or know how to recognize them, it has been a neuron firing dedicatedly and strongly to roses, which is exactly our definition for rose neuron(figure 3). \\
It's true that neurons don't take representing anything as their jobs from their perspective, yet their selectivity to certain entities makes them mental representations in effect from our perspective. Moreover, since the brain cannot access the outside world, it can only refer to the activity of these neurons with tight correlation to specific 'things' to 'see' what appears. In other words, it is the activation of these neurons instead the stimuli per se that directly determines what's in our inner world at the moment. If the rose neuron is inhibited or even damaged, then roses cannot appear in the mental world even they are projected to the retina. On the contrary, if the rose neuron is active, either spontaneously or by activated by electrodes, then the concept of rose will present in the mind. Failing to access the outside world, the brain has no way to judge whether roses really appear in the outside world. Thus, the brain's inaccessibility to the outside world not only doesn't deprive neurons of the ability of representing things in the environment, but actually endow them the authority for deciding what's present in our inner world. This ability means that these neurons are not only detectors, but they are indeed the inner representations that construct our inner world.\\
\begin{figure}[H]
    \centering
        \includegraphics[scale=.85]{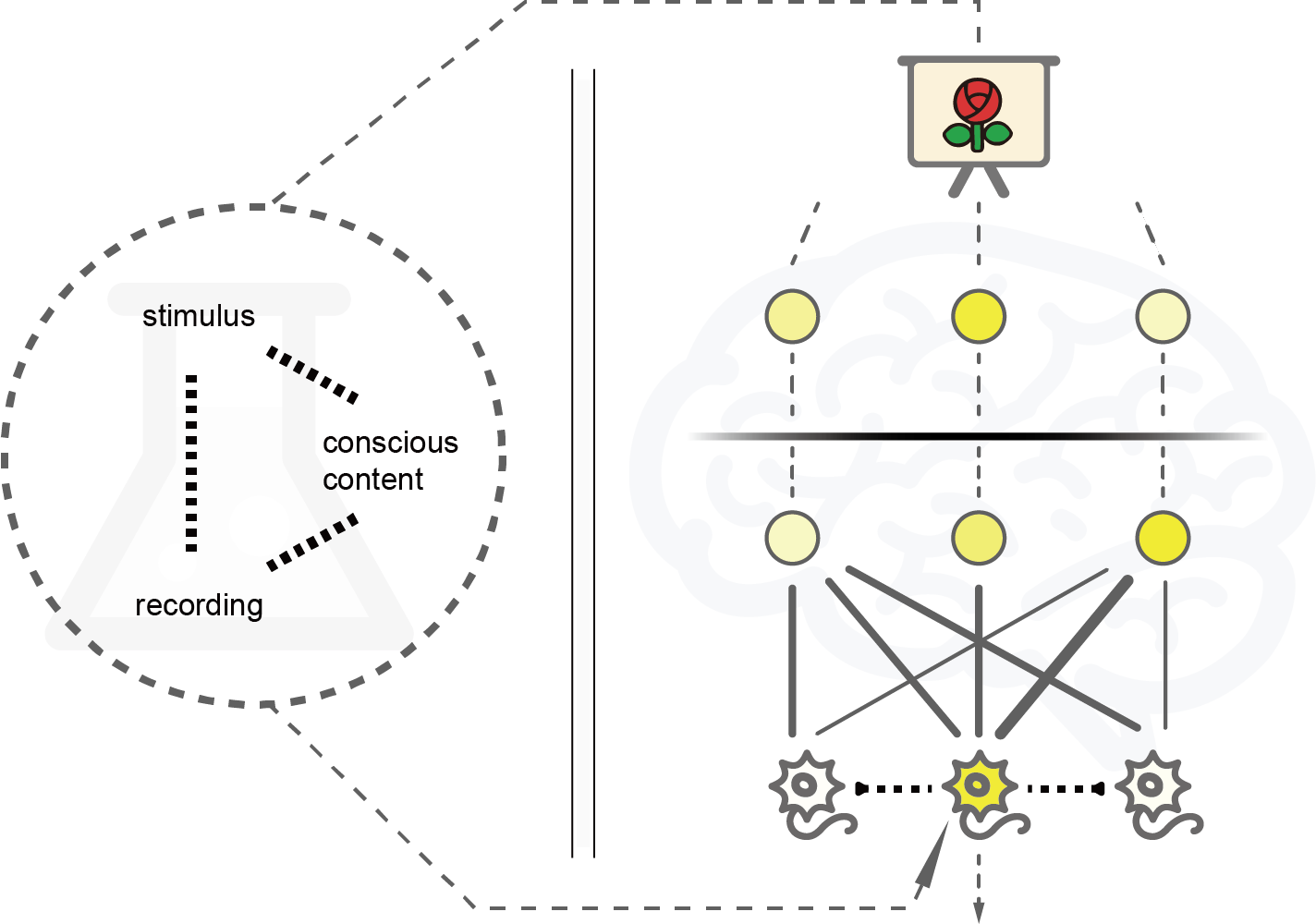}
    \caption{In the brain, a neuron wins the chance to fire to a activation pattern of its upstream neurons by making connections properly. Although this neuron don't know or see roses, it fires consistently and selectively to them. So in recording experiments, it can be found as rose neuron. In this way, two views of the brain, outside-in and inside-out, become compatible with each other.}
    \label{fig:3}
\end{figure}
Another eternal theme in the biological world is competition. Being the inner representation of something frequently present is not for free. Neurons have to participate in the fierce competition and only those firing strongly can finally occupy the conceptual space and consistently respond to the specific input patterns. This explains why it is reasonable to call the neuron firing strongly to roses a rose neuron instead of the neuron whose spike train is of a certain pattern(e.g.,0101 0010 0100 1111 0101 0011 0100 0101, the ASCII code for the word ROSE). The competition is inevitable. For one thing, it is also rooted in the selfishness of neurons because it's advantageous for a neuron to prevent other neurons from firing and taking shares. And for another thing, there is no way out. A neuron cannot retreat from the competition unilaterally, otherwise it will be knocked out by its peers that recruit inhibitory neurons. And neurons cannot stop the competition multilaterally. Since the competition as discussed can increase the sensory acuity of animals. Animals with inhibitory neurons shut up will hence have compromised sensory acuity and be eliminated by natural selection. \\
The fierce competition ensures that the needed neurons that can strongly respond to specific inputs will emerge. Again, although no neuron knows how to recognize roses or how to wire and arrange ion channels correctly to be responsible, the neuron that happens to wire appropriately and efficiently to respond strongly to roses can be selected out by the competition, which forces neuron to fire strongly, as the rose neuron automatically. This is just like creatures don't have to know what gene is fit for the environment, yet the right species with fit features and genomes will be selected out naturally.\\
As mentioned previously, computational models based on similar notions, Hebbian plasticity and competitive learning, have shown how neurons can become feature detectors in an unsupervised manner\cite{RN664}. Yet simulation is not equal to explanation. A problem with these models is that these two foundations are only arbitrarily imposed and cannot be proved solely from a computational perspective. But here, it is first proposed that these two assumptions can be plausible as they can help neurons to better survive in our brain. Besides, computational models will inevitably miss something because of the difference between computers and the brain. For example, competitive learning models cannot build associations among patterns, and hence other kinds of learning mechanisms are thought to be involved\cite{RN664}. Yet here, it is exactly the strong competition that forces neurons for associated concepts to establish reciprocal connections. These models usually require comparing neurons' excitation. Yet the thing is comparing 10 to 1 is as fast as 6 to 5 on computers, but can be much different in our brain as discussed ahead. The preclusion of the competition process as the result of enlarged excitation disparity will improve not only accuracy, but also efficiency. This cannot be immediately captured from a computational perspective. And the Theory of Information indicates a dense manner of coding will have a larger representational capacity which increases exponentially with the number of neurons while it can only increase linearly in a sparse scheme\cite{RN694}. Such a compromised encoding capacity seems a good reason for rejecting the idea of neuronal representations. However, from neurons’ perspective, this criticism is invalid because they only care about how to survive and will hence compete with each other to occupy territories alone even this competition will unfortunately lead to sparse coding and hence a compromised encoding capacity. This is another example that is contradictory to computational intuition since transistors don't take survival as their priority.

\subsection{Environmental reasons}
The environmental reason for a neuron to become specialized is that some entities appear frequently in our everyday life and create centers of gravity in conceptual space. Neurons that adjust their connections to be able to discharge at these entities can have advantages over their peers and survive in the brain. Therefore, these entities frequently present in one’s life will continually attract neurons to respond to them and largely determine what neurons one will have. One can have rose neurons only if roses are frequently present and can provide sufficient excitation for neurons representing them. And in turn, now that neurons actually live upon this frequently present entity, it provides another reason to call these neurons responding to roses rose neurons. \\
We cannot prepare all representations needed before birth. Hence neurons' chasing for the frequently present stimuli, which in turn determines what neuronal representations we will have, is important for allowing us to adapt to environments. There are lots of facts showing that responses of cortical neurons can be shaped by the environment. An induced orientation preference of primary visual cortex neurons of kittens reared in a controlled visual environment was reported long ago\cite{RN809}. Specifically, neurons of these cats reared in cylindrical chambers painted with black and white stripes at one orientation will preferentially respond to this selected orientation but weakly to others. In higher regions, it was found that one’s fusiform gyrus will respond more strongly to categories one is familiar with\cite{RN662}. These observations all support the notion that the abundance of inputs decides what neurons one will have. Importantly, this notion can further address some important theoretical issues.\\
Neuronal representation is criticized as vulnerable. As the joke goes, a person might suddenly forget his grandmother as the result of the accidental death of his cell grandmother coding. Yet it can be inferred from the previous discussion that this scheme, counter-intuitively, is robust and damage-resistant. The first point for its robustness is redundancy. Although the number of neurons coding for a certain entity is largely decreased by the strong competition, it will not be reduced to only one because the entity present frequently also brings more excitation and allows more neurons to survive upon it. The number of neurons representing a certain concept when the balance between excitability and competition is reached will be positively related to the familiarity of it\cite{RN829}. Hence, the loss of a few neurons won’t bring a disaster as we will have multiple grandmother cells as long as we are indeed intimate with our grandmother. It may be argued that this redundancy violates the principle of efficiency\cite{RN701}. But this is computational intuition again. Neurons only care about whether they can survive but not whether they are redundant. The strong competition has been trying to reduce redundancy, but if a neuron can still stand the competition and fire to the target, then there is no need to deny it. The second point is plasticity. Even in an extreme situation where all rose neurons are damaged, new rose neurons will emerge as long as the center of gravity still exists. And since there isn’t any special label attached to rose neurons, there is no reason to think these successor neurons will be any different. It is our surroundings that decide what neuronal representations will reside in our brain. That’s to say, if roses or our grandmothers are still frequently present in our life, we should never worry that we will fail to recognize them due to the accidental death of their dedicated neurons as new representational neurons will emerge continually.\\
Another important issue is what do the terms like entity, feature, concept, or category mean\cite{RN472,RN499}? These words have been extensively and casually used in previous literature and sections referring to the subjects of representations, but the question is what on earth is represented by neurons? Will only the general concept of rose have its corresponding unit? Or does each variety of rose have its dedicated unit? Or will we assign every rose we meet with a unit? This debate is usually related to the concern of combinational explosion. However, neurons themselves don’t know or care about the definition of these terms, but only care about whether they can earn enough excitation by following specific patterns. Again, the frequency of presence matters. If a specific rose has special meaning for someone and he or she sees this rose every day, then there will be enough reasons for neurons to follow the pattern elicited by this special rose and become its exclusive representational neurons. But other roses won’t have their dedicated neurons in his or her brain and they can only activate the neurons for the general concept of rose. For rose specialists, roses can create strong gravity in their brain which will then attract a large population of neurons. Consequently, different varieties of rose will all have their units. By contrast, for those who seldom see roses, the weak attraction of roses means only the general concept will be represented. The brain is neither a philosopher nor GOFAI. It doesn’t need that many presumptions to work. And there is either no need to worry about the combinatorial explosion because only frequently present entities can have their dedicated neuronal representations. If some entities fade away from our life, their gravity will be weak and their representation neurons will gradually turn to respond to other entities. So we will never run out of neurons.\\
A related issue is the granularity of neuronal representations. Due to the continuum nature of most things, it is hard to clarify how much interval is needed for two points on the continuum to become two things and represented by two different neurons. Should an oriented bar at 49° constitute a different thing compared with an oriented bar at 50 degrees\cite{RN499,RN868}? Again, it depends on the frequency of presence of the category. The logic behind is similar to that behind the fact that the rainforests have greater species diversity. For one thing, more neurons will be attracted by the familiar category initially and the subsequent stronger competition forces these neurons to be more selective. And for another thing, the frequent presence of this category allows neurons to survive upon very narrow subspaces of it. These two factors lead to fine-grained representations for familiar categories. Again, the granularity is not pre-defined, but is determined by the environment.

\section{How can neurons contribute to our cognition?}
Cognition is usually thought to be comprised of a series of procedures. Among these procedures, encoding is widely mentioned and discussed, whereas decoding is much overlooked. However, it should be reminded that encoding cannot be divorced from decoding\cite{RN592,RN546}, and information becomes information not when it is encoded, but when is decoded. It is argued that neuronal representation is impossible because our brain cannot decode information from their activities unless dualism is accepted. This is either not true because brain doesn't need a central decoder. Instead, each neuron, more specifically, settled neuron, can function both as an encoder and decoder. It decodes information from its upstream peers and carries specific meaning for its downstream peers. Therefore, the contribution of these settled neurons to our cognition should be discussed both as decoders and encoders.
\subsection{As decoders}
As decoders, the settled neurons extract the information embedded in the activation pattern of their upstream peers. It has been proposed and can be inferred from the previous discussion that settled neurons facing strong inhibition will respond to input patterns in a highly selective manner and can hence function as detectors. And their firing rate can indicate how much the input matches its targeted pattern. Moreover, their several properties ensure they can function as high-performance decoders.\\
It can be seen that settled neurons can function as efficient and rapid \emph{detectors}\cite{RN815,RN818,RN1547}. After the stabilization, neurons responding to different entities have concentrated their weights to their respective targets and set down their borders. When an input falls into a certain neuron’s territory, the explosive and dominant discharge of this neuron helps the detection to be done instantly and robustly even with much background noise, which mirrors our incredible perception ability for our familiar entities. The well-known ‘cocktail party effect’ is a good demonstration for it. Meanwhile, as its dominance helps to preclude the competition process, which consumes much time and energy and is full of uncertainty, we will feel effortless or even unconscious, and at the same time, finish the task accurately and rapidly when dealing with familiar entities represented by settled neurons. The 'pop-out effect' has shown us our almost automatic processing for familiar objects. In comparison, our perception for unfamiliar things will be tiring but unsatisfying as the result of the competition between unsettled neurons. This contrast mirrors our different language abilities. We can perceive fast speech in our own language effortlessly and accurately even in noisy environments, but listening comprehension when learning a new language can be demanding and difficult even the speech is slow and loud.\\
These settled neurons can also function as high-resolution \emph{classifiers}. The strong competition between neurons coding for familiar entities has forced them to differ with each other and respond to very specific entities respectively. Slightly different inputs will hence elicit distinct patterns at this level and can be readily distinguished, which allows us exquisite discrimination for familiar things. We can tell two alike friends but mix their faces up if they were just strangers to us. Similar pronunciations can be clearly discriminated for native speakers, but will sound the same for foreign people whose mother tongue doesn’t contain them. Here, the logic behind is quite plain. Possessing condensed excitation territories, the discharge of settled neurons can convey accurate information as they will only fire when a very specific input comes. By contrast, downstream neurons can only infer rough information from neurons with large territories and similar inputs cannot be discriminated now that they activate the same neuron. Such a trade-off between area and resolution is quite common in the brain. For example, in the primary visual cortex, the neurons responding to fovea own only narrow receptive fields, and accordingly, our central visual field has rather a high resolution. By contrast, neurons responding for the periphery have large receptive fields, and the precision of our peripheral vision is much worse. Similar things can be found in the motor system and somatosensory system. And things in associative regions are just the same.
\subsection{As encoders}
While being decoders means neurons can detect specific patterns in their upper level, being encoders means neurons can carry specific meanings and hence have particular value for their downstream neurons. Since settled neurons convey very specific meanings, they can underlie several critical cognitive functions.\\
It first functions to preserve information. It is argued that information is stored in trillions of synaptic weights\cite{RN472} instead of neurons. Yet although multiple observations show that experiences can modify synapses\cite{RN259,RN731,RN366,RN735}, this can only be partially correct because synapses can neither be accessed by other neurons nor exist independently. Only the discharge of neurons can be accessed by others and connections cannot break away from neurons. Arguing whether the information is stored in neurons or connections is like arguing whether the information is on the paper or the sentences. A more flexible way of thinking has to be taken. In this simplified model, the weight vector of a settled neuron encodes the coordinate of a specific point in the conceptual space. Analogically, these settled neurons possessing condensed territories can function as \emph{pins} nailed to specific spots in the conceptual space. Thanks to their precise preservation of the coordinates of specific points in the immense conceptual space, the specific pieces of information can be preserved and retrieved. By contrast, those nomadic neurons are not competent for preserving information as they cover large space and are highly unstable. Such a function of neurons can be related to our memory and be in agreement with our experience. For example, we can readily recall the appearance of people we are familiar with since familiar faces are encoded by settled neurons and thus can be easily retrieved. By contrast, we can hardly remember strangers’ appearance even we just met hours ago because we lack the pin preserving their spots in the conceptual space.\\
It has been advocated that neurons representing specific concepts can function as the \emph{building blocks} of our thoughts\cite{RN361,RN503}. Yet theoretically, comparing with bottom-up processing, top-down processing should be quite hard since it requires our brain the ability to initially and precisely activate particular neurons widely distributed to generate specific thoughts. This again emphasizes the importance of these settled neurons as they can serve as efficient and reliable proxies for the construction of certain information. Dealing with this small number of proxies is much easier and feasible. We may feel that top-down processing seems not so hard as it sounds to be. But this is actually because we have been thinking with prepared building blocks. We can readily combine concepts we are familiar with to composite new thoughts, but will have difficulty in thinking with unfamiliar entities. For a man who has only a rough concept of rose, it’s impossible for him to vividly envision a garden decorated with a variety of roses. Yet the same task can be easy for a gardener. By organizing these efficient and reliable neuronal representations in certain ways, one can effortlessly and continuously create new thoughts. Thus apparently, their such function can be related to our creativity.\\
These neurons further function as the \emph{basis} of learning for their upper layer. After stabilized, the interference between neurons is minimized. Hence, these neurons can now be described as orthogonal\cite{RN767,RN892} and can span a large and stable conceptual space, which is the prerequisite for the learning processing of their next layer. These settled neurons responding to inputs steadfastly and strongly will help their downstream to find centers of gravity readily so that learning can occur quickly\cite{RN755}. By contrast, if neurons at the lower level keep ranging and only fire occasionally and weakly to their targets, downstream neurons can never find a stable position to align with. At the behavior level, this mirrors our different learning capacities for things with different familiarity. Learning new concepts based on familiar elements can be much easier. For example, native speakers can readily learn new vocabularies but will feel hard to learn new words in a foreign language. This function is especially significant. It is usually criticized that we cannot prepare all neuronal representations needed. Yet this is exactly why we need neuronal representations because these settled neurons are crucial for creating the conceptual space, which in turn allows us fast learning for novel stimuli and adapt to novel environments rapidly.
\\
\\
Interestingly, accompanying the emergence of neuronal representations, many different dimensions of ability improve simultaneously. Native speakers can comprehend speech clearly and effortlessly. Elite boxers can judge distance precisely and rapidly on the court. Chess masters were reported to be able to read out latent information from the board and memorize positioning better than common people. Musicians are good at perceiving rhythms and new music flows out from their heads like waterfalls. This list can go much longer. Here, it is shown that how can settled neurons provide different functions and hence underlie various dimensions including efficiency, precision, and flexibility. Therefore, these seemingly different dimensions will improve simultaneously along with practice.\\
The popular word 'information' was almost never used in previous sections where neurons are only cells trying hard to survive. But in this section, it can be seen that the properties of cells surviving the fierce competition for entities frequently present make them into high-performance computation units even they themselves don’t intend to. Besides, their chasing for frequently present entities means we as a whole will selectively improve our performance for things we are dealing with daily or practicing intensively. By organizing these neurons hierarchically\cite{RN641,RN569}, we can process and produce complicated inputs and outputs efficiently\cite{RN738,RN799,RN590}. These improvements brought by these neurons concentrating on a small piece of territory actually serve as another reason for neuronal representations because they help the organism as a whole have a better chance of survival. 

\section{The power}
The discussion of the main topic of this paper has been completed, but the idea of neuronal representation and the thinking beyond it can be used to explore some other important problems. Moreover, what’s discussed here is something general, whether it can be applied to other domains can be an important criterion of its plausibility. If it's indeed valid, it must be compatible with many other phenomena and transferred to address other questions. Therefore, it is risky but also highly beneficial to discuss some seemingly irrelevant topics based on the previous discussion. 
\subsection{The question of neural coding}
The question of how our brain encodes concepts is the primary question of this paper. There have been several alternatives answers to this question among which localist coding and distributed coding have long been rivals. However, despite the lasting and heated debate between them, the definition of localist and distributed is quite inconsistent among researchers\cite{RN459,RN472,RN499}, which makes the previous debate somewhat pointless. Therefore, the first thing that should be done is to put previous definitions aside and clarify where the divergence actually occurs.\\
Our brain's hierarchical structure, redundancy in our brain, as well as the extensive association among neurons all imply that it's invalid to divide localist and distributed simply by whether neurons are repeatedly used or whether multiple neurons are activated by a given stimulus. Still, among existing literature, two questions are discriminative. First, can we read anything out directly from the activation of a single neuron? Or it is like a digit in information science and doesn’t convey any specific information when considered alone. Second, is there a leading neuron for a given entity at the tip of a processing hierarchy\cite{RN742}? Or such a leader is unnecessary and our brain represents complex inputs by forming dynamic assemblies\cite{RN706,RN647,RN748}. \\
These two questions can classify coding schemes into three types. And a scheme is localist only when the answers to both questions are yes. Take face recognition for example. If every face is encoded by a particular neural activation pattern while each neuron’s activation is context-dependent and cannot be interpreted alone, then it is doubtlessly distributed coding. If each neuron can represent a certain feature or variable in a context invariant manner, and a face is represented by a dynamic combination of neurons coding for different features, then it is still distributed scheme or population coding\cite{RN766,RN743} because there are only feature neurons but no face neurons. Lastly, if there does exist a neuron whose activation alone can be immediately related to a specific face, then it is localist coding even though there will inevitably be multiple other neurons responding to this face, especially neurons representing features at the lower level and it may respond to other faces as a result of co-activation.\\
\begin{figure}[H]
    \centering
    \includegraphics[scale=.85]{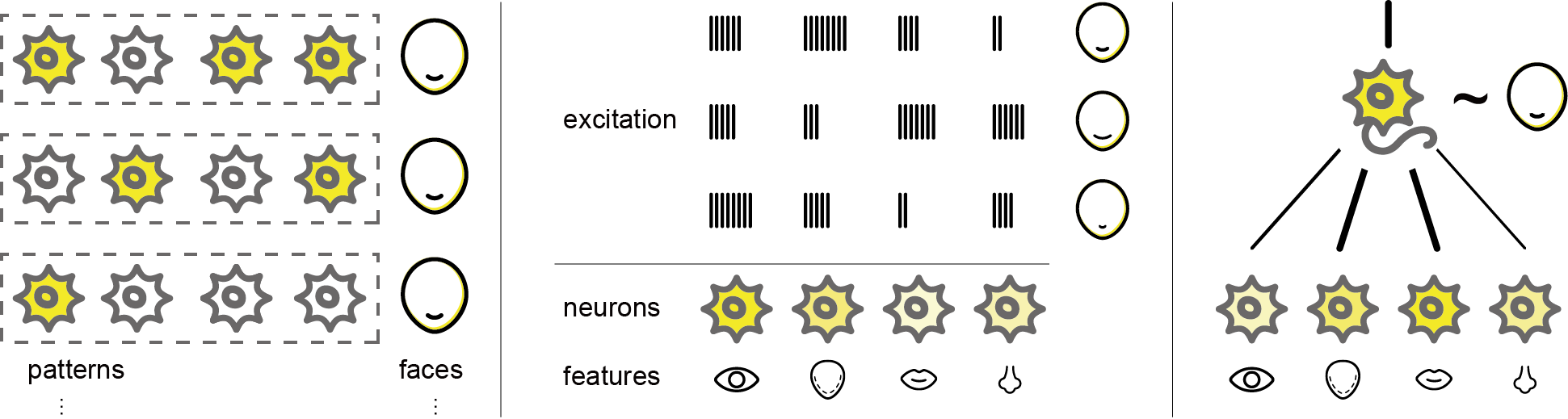}
    \caption{Distributed or localist coding.}
    \label{fig:4}
\end{figure}
The answer to the question of \emph{localist or distributed} is \emph{localist and distributed}. Our brain adopts both schemes and the familiarity matters\cite{RN499,RN708,RN829}. This can be easily inferred from the previous discussion. For those whom we deal with frequently, their strong gravity will attract multiple neurons and as discussed, there will be settled neurons coding for their identities and functioning as leading neurons. Their activation can be directly paired with certain people. But if a person is a stranger, his or her face will fall into the barren space and there won't be a dedicated neuron for him or her. In this situation, referring to any single neuron’s activation is meaningless. The brain has to take a population of feature neurons at the lower level to represent this person. In this narrative, the former is localist though a population of feature neurons at the lower level will be activated at the same time. And the latter is distributed though there can be neurons responding to specific facial features at the lower level. It is natural to assume that our brain must choose one from these two schemes especially when they seem exclusively the opposite. However, this is again only intuition and these two can incorporate with each other. \\
There have been experiments finding neurons encoding objects either in a distributed or localist manner. But now that our brain adopts both two schemes and they are actually compatible, finding evidence either for distributed coding or localist coding cannot immediately rule out the possibility of the other scheme. Importantly, it has been pointed out that the familiarity of materials used in experiments is the key factor\cite{RN537}. In experiments utilizing unfamiliar stimuli like strangers’ faces as materials, the results usually support the distributed coding, whereas evidence for localist coding comes from experiments using familiar stimuli. For example, in the famous experiment performed by Tsao and Chang\cite{RN486}, generated human faces, which are doubtlessly unfamiliar for the macaque subject, are used as stimuli and the result was unsurprisingly in favor of distributed coding. Interestingly, it shows each face cell serves as an axis and these neurons together span a linear high dimensional face space. Each face is coded by an ensemble of neurons or a point in this space and can be decoded precisely using linear regression. The face space found in this research is a perfect instance of conceptual space proposed previously and this experiment is a perfect demonstration for distributed or population coding in our brain. Conversely, evidence for a more localist scheme all came from experiments where familiar concepts like celebrities or famous architectures were used. And it is found that selective neuronal representations can be generated as the entities become familiar. This contrast between experiments using familiar or unfamiliar stimuli is in agreement with the previous discussion.\\
Outside the laboratory, our everyday experience also suggests that we adopt both two schemes but for different categories of entities. We have no difficulty in recognizing both strangers and acquaintances, but our capacity for these two categories of people is different. We can form a general impression of our friends' faces but have to turn to distinctive facial features to recognize strangers. We can always tell two friends even they look much alike, but will often mix two strangers up. The face of a friend can jump out from the crowd while a stranger’s face will not. We can run top-down perception to search for a friend readily but will have to continually refer to a photo to find a stranger. And we can recognize our long-missing friends but will fail to recognize a stranger met even just hours ago. All these facts showing our different capacities for strangers and acquaintances suggest that we adopt different coding schemes for these two categories, especially when the contrast between capacities echoes the earlier discussion of the contribution of settled neurons.\\
Criticisms of the idea of localist representation are usually on its incompetence facing the combinatorial explosion\cite{RN472,RN747}. Of course, we can’t assign everything we met with a dedicated neuron as the objects we would encounter will far outnumber the neurons we have. But it is permitted to allocate a portion of neurons for entities we deal with frequently and adopt the strategy of combination or distributed coding for others. For example, We cannot assign every number with a neuron but can use a combination strategy to \emph{generate} numbers infinitely. But for some special numbers, such as our phone numbers, we will generously take localist scheme so that they can be \emph{retrieved} readily.\\ 
An obvious question is now that population coding is already competent, why there should also be a localist scheme? There are two reasons. The first reason is that it is the inevitable result of neurons’ behavior as discussed earlier. The point here is these two schemes are not separated but continuous, and a shift from distributed coding to localist coding will gradually and naturally happen if an entity is repeatedly present\cite{RN853,RN636,RN529}. The second reason is from the function perspective. It has been correctly pointed out that localist coding allows rapid processing but is expensive whereas distributed coding sacrifices efficiency for flexibility\cite{RN708}. These two seemingly contradictory strategies are actually complementary and hence a collaboration of them can be highly efficient. While don’t have to worry about running out our neurons since entities appearing occasionally can be represented by the joint activation of neurons, we can take advantage of the efficiency provided by localist coding when dealing with entities we encounter daily. These entities are things important to us and it is hence worthwhile to trade flexibility for processing speed here. Interestingly, although the objections against neuronal representations are often from researchers with a background in information science, this is the same logic behind many designs in information science. For example, high-speed but expensive registers are used for frequent processing, and slow but cheap hard disks are used for storage. So taken together, this conclusion that our brain adopts both localist coding and distributed coding can both account for contradictory results of different experiments and other mental phenomena as well as be an efficient solution for neural coding.
\subsection{The influence from the past}
Several related theories, including active inference\cite{RN654}, predictive coding\cite{RN481}, and free energy principle\cite{RN487}, have been brought up in the last decades. Differently as they are stated, they are intrinsically consistent as they all emphasize that our brain’s processing information is also influenced by our past\cite{RN749,RN418}. By exploiting our experiences and performing certain optimizations, our brain will be able to reach an optimal state where we can react to current inputs efficiently. This proposal that our information processing capacity benefits from our experience is doubtlessly correct, yet the problem is that how can these conceptual models and algorithms can be neurally implemented in our brain.\\
The previous discussion about how neurons become specialized then serve as cognitive elements and how they adjust their connections with other cognitive nodes according to the synaptic history can provide insights for these questions. As discussed, it is neurons' strategy to wire more strongly with neurons that are more likely to fire simultaneously with them so that they can integrate enough excitation to overcome inhibition. While being a survival strategy for neurons, it helps the brain to build an inner probabilistic model where entities that are more related and hence more likely to present together will have stronger connections between their respective neuronal representations. Each neuron is not only a node coding for certain states but also an agent that autonomously adjusts connections according to probabilistic information extracted from synaptic history. This shows how our brain can utilize our experience. After that, stronger activation can be spread between these neurons wire more tightly, which endows our brain the predictive nature as the neurons for the more probable next states will receive stronger activation from the neuron for the current state and can therefore more easily be activated. This shows how can the past influence the future. Finally, in consistence with the previous discussion of the final stage of cortical development, it can be further proposed that the optimization our brain performs is to minimize the clashes and inference between neurons and to maximize its efficiency.\\
These proposals can reflect in our reactions under different conditions especially in scenarios where we feel surprised. Suppose a man who grew up on a farm and has only seen grey pigs running on the ground. Since he will be familiar with pigs as well as their color and acts, neurons coding for pigs, grayness, farms, and running respectively will form in his brain. Moreover, strong reciprocal connections will be established either directly or indirectly among these neuronal representations. When this man sees a grey running animal, strong reentrant signals will be spread to the pig neuron from the grayness neuron and the farm neuron. If the running animal is indeed a pig, this pig neuron will receive excitation from both upstream and downstream neurons and the integrated inputs will help it fire dominantly and abruptly, which helps this man to perceive the pig instantly and confidently even the detail is only vaguely checked\cite{RN752}. But for another man who doesn’t have the same background, it will take him seconds to confirm what is this running animal. However, when this farmer sees a pink pig flying above a factory, the ascending inputs of the perceived scenario will now diverge with the descending predictions. The grayness neuron receives an activation spread from the pig neuron while the pinkness neuron is activated by its afferent inputs. The moderate activation of different neurons at the same level will result in clashes and the consuming competition between neurons. And under this condition, this man as a whole will feel surprised or even uncomfortable. But the other man won't experience the same surprise. Human beings are known to be extremely sensitive to surprise, but what's less noticed is that the sensitivity is positively related to familiarity. It can be seen here that while the bidirectional connections between neurons coding for familiar and related entities facilitate our perception of them in a predictive manner, it also underlies our sensitivity to the change or mismatch in our familiar fields. \\
\begin{figure}[H]
    \centering
        \includegraphics[scale=.85]{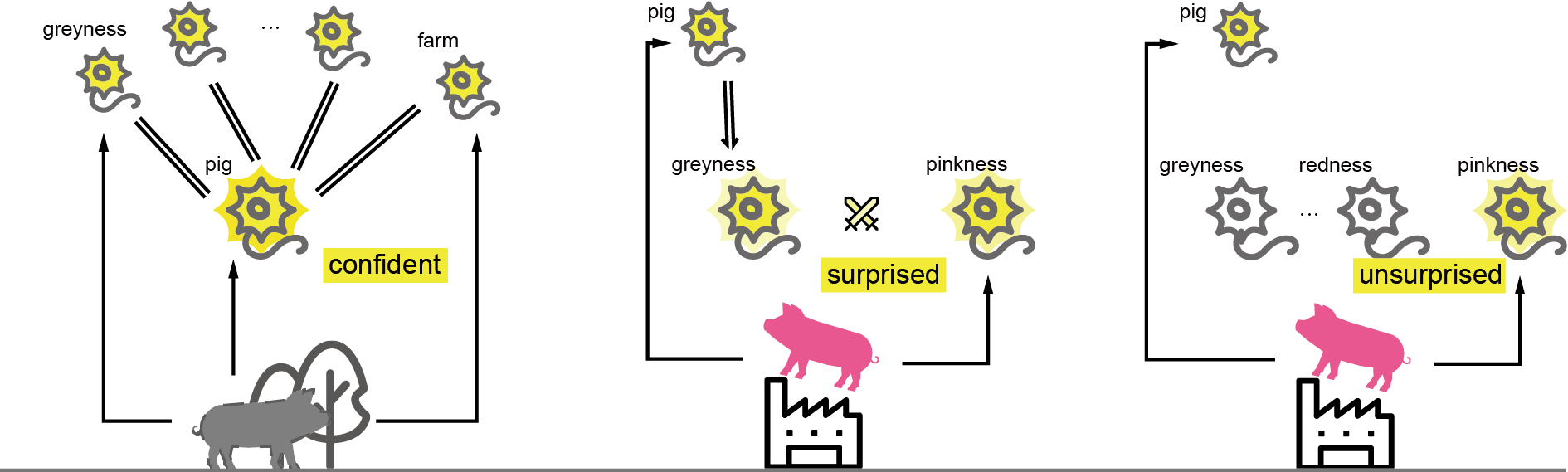}
    \caption{Left: integrating predictive feedback from neurons activated by other 'things' in the environment and the forward input, the pig neuron can fire dominantly. In this case, the perception is confident and rapid. Middle: the predictive feedback and forward input activate different color neurons respectively to a moderate level. The competition between them will be translated to the surprised reaction. Right: without the strong connections between color neurons and pig neuron, the other man won't be surprised by the pink pig as only pinkness neuron is activated.}
    \label{fig:5}
\end{figure}
The desired peaceful and stable state where the occurrence of clash and inference is minimized echoes with the previous discussion where neurons in the stabilization stage will set down their borders to reduce clashes between each other. Such consistency adds plausibility to the discussion here. And it is reasonable to regard peace and harmony as the goal of our brain because it also means efficiency for our brain. Although competition among neurons is critical for learning, it after all consumes much time and energy, which is clearly adverse for organisms. This disadvantage brought by competition may explain why we will feel somewhat unpleasant and uncomfortable when the perceived information is at odds with the predictive information. For example, it's annoying to use programs whose user interfaces are improperly designed and the feedback they provide is always at odds with our expectations based on our daily experience. And for people who are proficient in certain skills, sometimes little change in the setup will be intolerant and affect their performance, while people unpracticed in these skills even cannot notice the difference. Moreover, as the surprise results directly from the divergence of perceived information and prediction signals, minimizing the occurrence of clashes among neurons is equal to minimizing prediction error, which is in agreement with the proposals of theories mentioned ahead. \\
Taken together, it can be seen here that the neuronal representations can provide a biological implementation for computational theories mentioned and the interactions among these neurons can mirror their prediction or proposals, which in turn provides support for the idea of neuronal representations. Importantly, the discussion here precludes the need for an explicit model in our brain by replacing it with an implicit probability distribution that can emerge automatically as the result of neurons’ strategies for survival. Although neurons own no knowledge about conditional probability or the Bayesian theory and are only trying to integrate more excitation, their preference for choosing connection partners turns themselves into nodes coding for certain cognitive elements and our brain as a whole into a large Bayesian network. They don’t intend to predict or check the error. Yet the activation spread among neurons tightly connected will function as predictive signals and the clashes resulted from the divergence between inputs and predictive signals can indicate the occurrence of error. And although neurons have no idea about optimization, they will finally reach a state where the probability of error will be minimized as the clashes are avoided as much as possible. Such a relationship between the micro-level and macro-level is quite similar to what happens in the physical world where particles know nothing about distribution, yet the gas as a whole can be described by probability distribution at the macro level magically. Such an idea, while proved useful in physics, can be instructive in understanding the relationship between our brain and neurons within it as well.\\

\subsection{The problem of binding and perception}
The binding problem\cite{RN784,RN560} is an important theoretic problem in both psychology and neuroscience. Two theories, Feature Integration Theory (FIT)\cite{RN764,RN763,RN559} and Binding by Synchrony (BBS)\cite{RN508,RN420,RN702} were brought up to account for the binding process. However, while both theories receive supports from considerable empirical works, they are contradictory to a certain degree. FIT emphasizes the role of attention and the involvement of higher-order regions, whereas BBS claims that binding is an automatic process arising from low-level neural dynamics\cite{RN751}. Meanwhile, both theories are criticized for the vague description of their core mechanisms. FIT is criticized for its vague treatment of attention\cite{RN506}. And BBS is criticized for the absence of detailing how binding is computed\cite{RN789,RN507}. Hence, though offering lots of insights, they are still halfway to the final answer.\\
The main topic of this paper, neuronal representations are thought to play an important role in binding\cite{RN789,RN702,RN793}. In different regions at lower levels, neurons responding to features of coming sensory inputs will be activated so that these features can be detected. Such fast detection is in agreement with the description of the registration stage\cite{RN762} and the synchronized activation of these feature detectors are the basis of BBS. Yet the questions are how can synchronized activation of neurons widely distributed lead to binding and what are the roles of attention and higher regions. Here, I will answer these questions within the framework proposed in the current paper.\\
For an object only occasionally present like a yellow triangle, though there can be prepared neurons for the yellowness and triangles due to their frequent presence, it's not likely that there is also a prepared yellow triangle neuron in one's brain. When a man sees a yellow triangle, yellowness neuron and triangle neuron at lower levels will fire strongly together, which will create a center of gravity in the conceptual space of higher-level neurons. Yet this point is not within any settled neuron's territory and no neuron at higher levels can strongly respond to it. To create a neuron representing this instance for further processing, attention has to play its role. It can be speculated that attention here is to temporarily disturb the established excitatory-inhibitory balance through neuromodulation\cite{RN577,RN718,RN547,RN713,RN779,RN360} and allow neurons stuck by inhibition to move around on the conceptual space to search for the yellow triangle point somewhere by temporarily adjusting its synaptic strengths\cite{RN792,RN795}. They will start their searching because the point created by concurrent and convergent inputs is attractive to them.\\
While attention functions as facilitation instead of glue, neurons at upper levels that are set free by attention and successfully spotting the concurrent inputs of yellowness neuron and triangle neuron in the conceptual space can function as the glue. By adjusting its connections with these two neurons, this neuron will become a temporary yellow triangle neuron. Then, it will enhance its feedback connections with these two upstream neurons. In this way, they can form a temporary reentrant system where the temporary yellow triangle neuron can function as a router to help to synchronize the activities of neurons within this system\cite{RN506,RN705,RN714,RN1534}. The resonance among these neurons can help these neurons to overcome the inhibition they are facing while the rest non-grouped neurons will be inhibited\cite{RN708,RN1534}. Therefore, the synchronized activation of neurons is the result of binding in a sense. If two things are not seen as from one object, then the coherence of neurons will not increase despite their increasing firing rates. But this synchronized activation in turn becomes the requirement of binding because the higher region neuron has to receive spikes arriving within the narrow window to keep firing.\\
Taken together, binding can be described as the process where neurons at higher regions search for the point in the conceptual space created by the synchronized activation of neurons for the componential features at lower levels with the facilitation from attention. The synchronized activation of feature neurons at lower levels, which is the core of BBS but seems to be much dismissed in FIT, is crucial for binding because it creates the point in the conceptual space for higher region neurons to search for. Yet the synchrony alone cannot achieve binding immediately. Patients with Balint’s syndrome or integrative agnosia\cite{RN797} can still detect features but fail to bind them together. This cannot be explained if synchrony is everything for binding. Attention is apparently crucial for binding, yet while there has been much research showing its function of modulating neurons’ activities, there isn’t any evidence suggesting it can function as glue. A more plausible account for its role is to assist higher-level neurons to find the points. For one thing, it modulates the activities of higher-order neurons to enable them to move around, which mirrors the fact that while basic features can pop out and be processed subconsciously, we will feel that we have to be attentive to recognize complex inputs. For another thing, when the binding becomes hard due to the distractors, spatial information and selective attention will be needed\cite{RN692,RN716,RN759}, which may make the point created by one object salient through selective enhancement so that higher-order neurons can find it in the conceptual space easily. These two different roles of attention are in agreement with the recent view that attention is not unitary but should be categorized as internal attention and external attention\cite{RN805}. Finally, it is the neuron receiving convergent inputs that binds features together. \\
\begin{figure}[H]
    \centering
        \includegraphics[scale=.85]{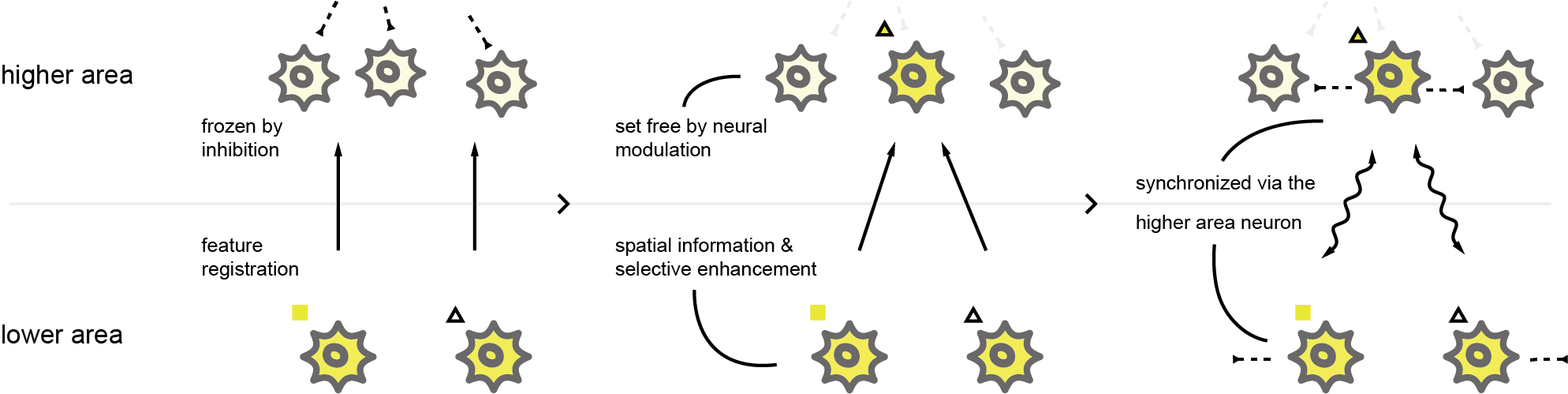}
    \caption{Three stages of binding.}
    \label{fig:6}
\end{figure}
It is argued that it can’t be neurons at upper levels that function to bind features for the concern of combinatorial explosion\cite{RN782,RN702}. However, as analyzed earlier, such a worry is unnecessary. Besides, there are more reasons for the involvement of higher regions. First, while the bidirectional connections between neurons at lower levels may function for synchronization, such a scheme is uneconomical in terms of the cost for massive connections. The scheme where higher regions receiving convergent inputs function as hubs to connect neurons widely distributed\cite{RN293,RN322,RN326} can be much more efficient\cite{RN284,RN338}. Second, the relatively small receptive fields of lower-level neurons cannot account for our capacity of perceiving objects on a large scale whereas the larger receptive field of neurons at higher levels can offer a reasonable explanation. And more importantly, such a worry about combinatorial explosion reflects the neglect of the brain’s plasticity. The neuron at the upper level is only a temporal yellow triangle neuron and can be reused for future binding after the disappearance of the yellow triangle. \\
Ironically, the definition of feature is unclear despite its wide use. Although we usually regard colors, orientations, and geometries as features and their combinations as complex stimuli, such categorization is quite arbitrary and ambiguous. Letters are consisted of geometries but are classified into features instead of complex stimuli. This ambiguity makes the discussion of the binding problem somewhat baseless. Yet under this framework, features can be defined as entities that are already assigned with settled neurons due to their frequent presence while complex stimuli don’t have dedicated correspondence. It is the absence of prepared neuronal representations that urges the need for a binding process. For Chinese people, characters are basic features though they are objectively complex. It can be effortless for the Chinese to recognize characters. But for other language speakers, characters are complex inputs made up of a bunch of lines and it takes much attention to bind these lines together. It takes Chinese attention to bind characters with their motions since there are no prepared neurons for these combinations. Yet it is still an easy job thanks to the prepared conceptual space. But this task will be a nightmare for foreign people due to the lack of character neurons. That’s to say, although the objective complexity of different stimuli varies, there isn't any prior rule deciding whether an entity is a feature or a complex stimulus. And it is therefore baseless to presuppose that some entities can be represented by individual neurons while others by the joint activation of a population of neurons. The difference between features and complex stimuli is that the former have gone through thousands of times of binding and hence are represented by settled neurons that have already tightly bound with their upstream neurons while the latter have to recruit neurons to temporarily bind feature neurons together. Repeated exposure to complex stimuli will turn these temps into regular workers, which mirrors the facts that the practice will facilitate the processing and even turn the processing automatic\cite{RN801,RN806} and echoes with the previous discussion about the emergence and cognitive significance of neuronal representations.\\

\section{Discussion and Conclusion}
Although humans have long known that the brain is the palace for the mind and researchers from different fields are all enthusiastic in figuring out the relationship between biology structure and mental phenomena, the gap between brain and mind seems to get never closer. This failure urges an inspection of the ways we have adopted. \\
Researchers from different backgrounds adopt different methods. The first team consists of theoretical neuroscientists who resolve to search for the answer in computational models. Their works have indeed promoted the development of artificial intelligence\cite{RN270}. Yet this is after all an indirect way for understanding the brain. And these models adopt too many arbitrarily imposed rules, especially many of them can be described as “fly in the face of constraints provided by the operations made available by neural hardware”\cite{RN462}. According to Marr, computational neuroscience is dealing with the computational level but not the implementational level. But if this is the proposal, the gap between brain and mind will be broader unsurprisingly. Another contingent is led by biological neuroscientists including several Nobel laureates. Thanks to their intelligence and diligence, numerous biological details in the nervous system at the molecular level have been unveiled, which remarkably enhances our understandings of our brain. But on the other hand, they seem to pay too much attention to details. The emergent property of our brain and mind implies that the extensive interaction among billions of neurons is the key to our questions. It seems the deeper we dive into details, the broader the gap will be. The third group is comprised of psychologists and cognitive neuroscientists\cite{RN738}. Inherited terms from psychology, they are keen to find the neural correspondence of them in our brain, which could be regions, circuits, or certain brain activities\cite{RN595,RN737,RN570}. Since our mind emerges from the activities of our brain, we can always find something related. Yet it is this emergent property that implies that this is an invalid way. Different levels have different logic and hence should have different vocabularies. We can only use momentum but not pressure or temperature for molecules’ movement. Likewise, neurons have their own logic and the problem here is not about where on earth these terms come from\cite{RN594}, but they are invalid for understanding the brain and neurons' logic. Rudely imposing terms on the brain won’t help to truly understand the brain or close the gap. Briefly speaking, in these three paradigms, neurons are respectively fictional, isolated, and subordinate.\\ 
In this paper, I introduce a novel way to treat neurons where I highlight neurons' essence as living cells instead of immortal operators and ask how can they survive in the brain. Then I demonstrate how the selectional presence upon neurons’ excitability and neuron’s strategies for survival together drive them to strongly, selectively, and dedicatedly to input patterns frequently present at the micro-level, which ensures them to become neuronal representations for familiar entities in practice at the macro-level. The introduced concept of excitation territory, which can be taken as neurons’ habitats, is useful here. Compared with its correspondence tuning curve, this concept allows to not to describe the electrophysiological profiles of individual neurons, but also the interaction among them. Neurons' contending for habitats to survive is the thread running through the whole process of learning. \\
Although this is a paper in defense of neuronal representations and hence localist coding, it also supports the idea of population coding. The introduced concept of conceptual space links these two schemes together. Spanned by neurons representing different concepts, this space allows both representing numerous possible entities in a distributed way and forming neuronal representations for familiar entities rapidly. It can be seen that these two opposite schemes of coding can be compatible and complementary. Their collaboration yields a coding scheme with both efficiency for familiar entities and flexibility for unfamiliar entities. Their division of labor is in agreement with the demonstration of the emergence of neuronal representations and a range of cognitive phenomena. Furthermore, as the position of a neuron in the conceptual space determined by its connections defines its role in information process, this concept finds a link between the biology and cognition. Thus, starting from defending neuronal representations, this paper ends up with providing a theory about how our brain might work.\\
As the title of this paper indicates, a large part of the discussion here is from the perspective of neurons. Today, almost all discussion is only from our perspective. Common questions are how neurons can support our information processing or what are the functions of various neuronal mechanisms. However, the first problem is the mismatch between the intention, which is ours, and the action, which is neurons’. The question that what do these phenomena mean for neurons is rarely asked. In the first half of this paper, by putting ourselves in neurons' shoes and asking how their actions accord with their goal of survival, I have shown how can we provide reasonable explanations for a range of observations in the brain and a complete narrative for the developing and learning process. Neurons' behavior is usually as inconsistent as British foreign policy, but this notion that neurons are acting to survive can serve as the polar star for interpreting and analyzing this variance. Then the second problem is that while researchers take a straightforward way which goes like now that our brain is a “computer’, neurons must be computational elements, nature is usually not so direct. Differently, I take a roundabout way where I avoid discussing neurons' computational properties at first. Then in the second half of this paper, it's shown that while neurons only care about how to gather more excitation, their action taken to achieve so underlie our various cognitive capacities magically. An interesting point here is while different cognitive capabilities like memory, imagination, and perception are distinctive topics from psychologists’ view, they are highly related when viewed from neurons perspective, just like temperature and pressure are completely different, but are related when viewed from the micro-level. So taken together, this paper also provides new threads for considering neurons' behavior and unifying the neuron level with the cognition level\cite{RN862}. \\
While simplifying the situation for the sake of generality, this paper doesn’t claim that the brain is that easy. It shouldn't be taken as neurons in different regions and species are all the same or the brain is a blank slate and learns everything from the environment. We surely cannot teach animals our language simply by exposing them to our language environment. The exact mechanisms underlying our incredible capacities are crucial and still to be elucidated. But on the other hand, what's discussed here is something general. Just like the concision of the Theory of Evolution is in no contradiction to the complication of molecular biology while the former actually helps to discover and understand the latter, it is possible and even necessary to first obtain a rough and general principle behind complicated observations. Science is to a certain degree similar to painting. It's unwise to dive deep into details at an early stage whereas starting from a clean outline and then adding in details progressively is a more promising strategy. Besides, it is beneficial to separate the discussion of abstract principles from that of exact mechanisms. An effective general rule can not only incorporate mounting details easily, but also be instructive for understanding details. As a good demonstration, terms like evolution, competition, and adaptation don’t directly describe how features can be a result of gene expression and be passed to offspring, but they turn out to be compatible with our later findings at the molecular level and more importantly, provide a powerful context for discussing these observations. Sitting at another level, the general principles can help to explain and organize the existing data and predict what can be expected, which are exactly what the science is aiming for. We may never know the exact wiring pattern of the brain due to its extreme complexity, diversity, and flexibility of the brain, yet we can still foresee what will happen in our brain with the effective framework, just like we don't have to dive into molecular levels to predict what will happen in the ecosystems. Nevertheless, the painting can be finished only after tons of details have been added to the rough sketch. This paper is only a beginning and a little window, numerous questions are to be answered.\\
A key point of this paper, as repeatedly appear through the previous discussion, is that neurons and our minds exhibit a parallel relationship. While we are representing the entities and their associations with the help of neurons and connections, these cells inside our brain don’t intend to present anything. They are selfish cells and only care about how to fire more and survive. They wire with each other only to gain more excitation, yet their wiring strategies magically turn our brain into a 'computer'. The stories of us and that of neurons inside our brain seem like two totally different stories happening in parallel spaces where we and neurons are just doing respective business independently and autonomously. However, two seemingly divergent storylines magically converge. Such a relationship between the micro- and the macro-level is commonplace in other subjects. For example, molecules constituting the gas in the cylinder are just there moving around and colliding with each other randomly and autonomously. However, their meaningless movements allow the gas as a whole to be described by the ideal gas law and moves the piston. Again, seemingly divergent storylines converge. \\
Scientists actually have no difficulty in dealing with such a strange relationship between levels. It won’t be claimed that the GAS can control the movement of constituent molecules to push the piston or read the movements of molecules to decide its temperature. However, when the subject is ourselves, we have repeatedly made these mistakes that could have been easily avoided. Our rooted belief makes it hard for us to envision the same parallel relationship between us and our neurons. We would rather believe that neurons in our brain are obediently and passive servants than accept that they are also autonomous and have their own intentionality\cite{fitch2008nano}. This idea that the beings of neurons are in parallel with the beings of us will radically challenge our current beliefs and raise serious philosophical questions: \emph{Who are we then?}\\
But this anyway a promising way for solving the problem we are currently facing. Humans have used the same strategy to unify our knowledge at different levels. But our reluctance to apply this powerful strategy to ourselves results in a breakpoint between biology and psychology, which splits science into hard and soft science and hinders the ambitious goal of consilience\cite{RN826}. Philosophy again gets in the way. Here, as just mentioned and demonstrated, I still suggest taking distinct ways of thinking at different levels. At the lower level or the neural level, we have to respect the fact that neurons are organisms and avoid concepts from cognitive science or psychology. At the higher level or behavioral level, we then move our eyes off from individual neurons but ask how our cognition emerges from their interaction. I have shown how divergent storylines at different levels can converge magically and addressed several important questions. And its power won't be confined to the topics discussed in this paper, but can potentially be applied to address a number of other problems of our brain and mind. Humans have used the same strategy to unify other subjects. And promisingly it can bridge the gap between neuroscience and psychology, more importantly, the gap between hard science and soft science as well.     
\medskip

\bibliography{main.bib} 

\begin{thebibliography}{100}

\bibitem{RN448}
J.~S. Bowers, ``On the biological plausibility of grandmother cells:
  implications for neural network theories in psychology and neuroscience,''
  {\em Psychol Rev}, vol.~116, no.~1, pp.~220--51, 2009.

\bibitem{RN465}
J.~S. Bowers, ``Grandmother cells and localist representations: a review of
  current thinking,'' {\em Language, Cognition and Neuroscience}, vol.~32,
  no.~3, pp.~257--273, 2017.

\bibitem{RN503}
R.~Q. Quiroga, ``Concept cells: the building blocks of declarative memory
  functions,'' {\em Nat Rev Neurosci}, vol.~13, no.~8, pp.~587--97, 2012.

\bibitem{RN640}
R.~Q. Quiroga, ``Gnostic cells in the 21st century,'' {\em Acta Neurobiol Exp
  (Wars)}, vol.~73, no.~4, pp.~463--71, 2013.

\bibitem{RN460}
H.~B. Barlow, ``Single units and sensation: a neuron doctrine for perceptual
  psychology?,'' {\em Perception}, vol.~1, no.~4, pp.~371--94, 1972.

\bibitem{RN461}
C.~G. Gross, ``Genealogy of the "grandmother cell",'' {\em Neuroscientist},
  vol.~8, no.~5, pp.~512--8, 2002.

\bibitem{RN501}
A.~S. Barwich, ``The value of failure in science: The story of grandmother
  cells in neuroscience,'' {\em Front Neurosci}, vol.~13, p.~1121, 2019.

\bibitem{RN833}
K.~A.~C. Martin, ``A brief-history of the feature detector,'' {\em Cerebral
  Cortex}, vol.~4, no.~1, pp.~1--7, 1994.

\bibitem{RN860}
D.~Rose, ``Some reflections on (or by?) grandmother cells,'' {\em Perception},
  vol.~25, no.~8, pp.~881--6, 1996.

\bibitem{RN542}
D.~H. Hubel and T.~N. Wiesel, ``Receptive fields, binocular interaction and
  functional architecture in the cat's visual cortex,'' {\em J Physiol},
  vol.~160, pp.~106--54, 1962.

\bibitem{RN569}
D.~C. Van~Essen, C.~H. Anderson, and D.~J. Felleman, ``Information processing
  in the primate visual system: an integrated systems perspective,'' {\em
  Science}, vol.~255, no.~5043, pp.~419--23, 1992.

\bibitem{RN839}
K.~Grill-Spector and R.~Malach, ``The human visual cortex,'' {\em Annu Rev
  Neurosci}, vol.~27, pp.~649--77, 2004.

\bibitem{RN841}
C.~G. Gross, ``Representation of visual stimuli in inferior temporal cortex,''
  {\em Philos Trans R Soc Lond B Biol Sci}, vol.~335, no.~1273, pp.~3--10,
  1992.

\bibitem{RN647}
N.~K. Logothetis, J.~Pauls, and T.~Poggio, ``Shape representation in the
  inferior temporal cortex of monkeys,'' {\em Curr Biol}, vol.~5, no.~5,
  pp.~552--63, 1995.

\bibitem{RN469}
R.~Q. Quiroga, L.~Reddy, G.~Kreiman, C.~Koch, and I.~Fried, ``Invariant visual
  representation by single neurons in the human brain,'' {\em Nature},
  vol.~435, no.~7045, pp.~1102--7, 2005.

\bibitem{RN636}
R.~Quian~Quiroga, A.~Kraskov, C.~Koch, and I.~Fried, ``Explicit encoding of
  multimodal percepts by single neurons in the human brain,'' {\em Curr Biol},
  vol.~19, no.~15, pp.~1308--13, 2009.

\bibitem{RN546}
M.~György~Buzsáki, {\em The brain from inside out}.
\newblock Oxford University Press, 2019.

\bibitem{RN651}
T.~D. Goode, K.~Z. Tanaka, A.~Sahay, and T.~J. McHugh, ``An integrated index:
  Engrams, place cells, and hippocampal memory,'' {\em Neuron}, vol.~107,
  no.~5, pp.~805--820, 2020.

\bibitem{RN258}
S.~Tonegawa, X.~Liu, S.~Ramirez, and R.~Redondo, ``Memory engram cells have
  come of age,'' {\em Neuron}, vol.~87, no.~5, pp.~918--31, 2015.

\bibitem{RN581}
E.~Kohler, C.~Keysers, M.~A. Umilta, L.~Fogassi, V.~Gallese, and G.~Rizzolatti,
  ``Hearing sounds, understanding actions: action representation in mirror
  neurons,'' {\em Science}, vol.~297, no.~5582, pp.~846--8, 2002.

\bibitem{RN582}
R.~Mukamel, A.~D. Ekstrom, J.~Kaplan, M.~Iacoboni, and I.~Fried,
  ``Single-neuron responses in humans during execution and observation of
  actions,'' {\em Curr Biol}, vol.~20, no.~8, pp.~750--6, 2010.

\bibitem{RN830}
R.~Q. Quiroga, R.~Mukamel, E.~A. Isham, R.~Malach, and I.~Fried, ``Human
  single-neuron responses at the threshold of conscious recognition,'' {\em
  Proceedings of the National Academy of Sciences of the United States of
  America}, vol.~105, no.~9, pp.~3599--3604, 2008.

\bibitem{RN1633}
H.~Gelbard-Sagiv, R.~Mukamel, M.~Harel, R.~Malach, and I.~Fried, ``Internally
  generated reactivation of single neurons in human hippocampus during free
  recall,'' {\em Science}, vol.~322, no.~5898, pp.~96--101, 2008.

\bibitem{RN457}
J.~Feldman and D.~Ballard, ``Connectionist models and their properties,'' {\em
  Cognitive Science}, vol.~6, no.~3, pp.~205--254, 1982.

\bibitem{RN587}
J.~Grainger and A.~M. Jacobs, {\em On localist connectionism and psychological
  science}, pp.~11--48.
\newblock Psychology Press, 2013.

\bibitem{RN755}
M.~J. Ison, R.~Quian~Quiroga, and I.~Fried, ``Rapid encoding of new memories by
  individual neurons in the human brain,'' {\em Neuron}, vol.~87, no.~1,
  pp.~220--30, 2015.

\bibitem{RN459}
M.~Page, ``Connectionist modelling in psychology: a localist manifesto,'' {\em
  Behav Brain Sci}, vol.~23, no.~4, pp.~443--67; discussion 467--512, 2000.

\bibitem{RN748}
R.~Quian~Quiroga, ``No pattern separation in the human hippocampus,'' {\em
  Trends Cogn Sci}, vol.~24, no.~12, pp.~994--1007, 2020.

\bibitem{RN545}
D.~O. Hebb, {\em The organization of behavior: A neuropsychological theory}.
\newblock Psychology Press, 2005.

\bibitem{RN554}
V.~Castellucci, H.~Pinsker, I.~Kupfermann, and E.~R. Kandel, ``Neuronal
  mechanisms of habituation and dishabituation of the gill-withdrawal reflex in
  aplysia,'' {\em Science}, vol.~167, no.~3926, pp.~1745--1748, 1970.

\bibitem{RN810}
X.~Liu, S.~Ramirez, P.~T. Pang, C.~B. Puryear, A.~Govindarajan, K.~Deisseroth,
  and S.~Tonegawa, ``Optogenetic stimulation of a hippocampal engram activates
  fear memory recall,'' {\em Nature}, vol.~484, no.~7394, pp.~381--5, 2012.

\bibitem{RN525}
X.~Liu, S.~Ramirez, R.~L. Redondo, and S.~Tonegawa, ``Identification and
  manipulation of memory engram cells,'' {\em Cold Spring Harb Symp Quant
  Biol}, vol.~79, pp.~59--65, 2014.

\bibitem{RN365}
G.~Vetere, L.~M. Tran, S.~Moberg, P.~E. Steadman, L.~Restivo, F.~G. Morrison,
  K.~J. Ressler, S.~A. Josselyn, and P.~W. Frankland, ``Memory formation in the
  absence of experience,'' {\em Nat Neurosci}, vol.~22, no.~6, pp.~933--940,
  2019.

\bibitem{RN549}
D.~L. Schacter, I.~G. Dobbins, and D.~M. Schnyer, ``Specificity of priming: a
  cognitive neuroscience perspective,'' {\em Nat Rev Neurosci}, vol.~5, no.~11,
  pp.~853--62, 2004.

\bibitem{RN553}
H.~Otgaar, M.~L. Howe, P.~Muris, and H.~Merckelbach, ``Associative activation
  as a mechanism underlying false memory formation,'' {\em Clinical
  Psychological Science}, vol.~7, no.~2, pp.~191--195, 2018.

\bibitem{RN361}
S.~M. Frankland and J.~D. Greene, ``Concepts and compositionality: In search of
  the brain's language of thought,'' {\em Annu Rev Psychol}, vol.~71,
  pp.~273--303, 2020.

\bibitem{RN539}
B.~B. Averbeck, P.~E. Latham, and A.~Pouget, ``Neural correlations, population
  coding and computation,'' {\em Nat Rev Neurosci}, vol.~7, no.~5, pp.~358--66,
  2006.

\bibitem{RN556}
G.~E. Hinton, {\em Connectionist learning procedures}, pp.~555--610.
\newblock Elsevier, 1990.

\bibitem{RN538}
J.~L. McClelland, D.~E. Rumelhart, and P.~R. Group, {\em Parallel distributed
  processing}, vol.~2.
\newblock MIT press Cambridge, MA, 1986.

\bibitem{RN694}
A.~Spanne and H.~Jorntell, ``Questioning the role of sparse coding in the
  brain,'' {\em Trends Neurosci}, vol.~38, no.~7, pp.~417--27, 2015.

\bibitem{RN543}
R.~Desimone, ``Face-selective cells in the temporal cortex of monkeys,'' {\em J
  Cogn Neurosci}, vol.~3, no.~1, pp.~1--8, 1991.

\bibitem{RN846}
R.~Q. Quiroga, I.~Fried, and C.~Koch, ``Brain cells for grandmother,'' {\em
  Scientific American}, vol.~308, no.~2, pp.~30--35, 2013.

\bibitem{RN564}
G.~A. Carpenter and S.~Grossberg, ``Art 2: Self-organization of stable category
  recognition codes for analog input patterns,'' {\em Applied optics}, vol.~26,
  no.~23, pp.~4919--4930, 1987.

\bibitem{RN498}
S.~Grossberg, ``Developmental designs and adult functions of cortical maps in
  multiple modalities: Perception, attention, navigation, numbers, streaming,
  speech, and cognition,'' {\em Front Neuroinform}, vol.~14, p.~4, 2020.

\bibitem{RN562}
K.~Fukushima and S.~Miyake, {\em Neocognitron: A self-organizing neural network
  model for a mechanism of visual pattern recognition}, pp.~267--285.
\newblock Springer, 1982.

\bibitem{RN491}
C.~von~der Malsburg, ``Self-organization of orientation sensitive cells in the
  striate cortex,'' {\em Kybernetik}, vol.~14, no.~2, pp.~85--100, 1973.

\bibitem{RN664}
D.~Rumelhart and D.~Zipser, ``Feature discovery by competitive learning,'' {\em
  Cognitive Science}, vol.~9, no.~1, pp.~75--112, 1985.

\bibitem{RN1375}
D.~Marr, {\em Vision: A computational investigation into the human
  representation and processing of visual information}.
\newblock 1982.

\bibitem{RN464}
E.~Thomas and R.~French, ``Grandmother cells: much ado about nothing,'' {\em
  Language, Cognition and Neuroscience}, vol.~32, no.~3, pp.~342--349, 2016.

\bibitem{RN844}
T.~C. Schelling, {\em Micromotives and macrobehavior}.
\newblock WW Norton \& Company, 2006.

\bibitem{RN811}
S.~Shoham, D.~H. O'Connor, and R.~Segev, ``How silent is the brain: is there a
  "dark matter" problem in neuroscience?,'' {\em J Comp Physiol A Neuroethol
  Sens Neural Behav Physiol}, vol.~192, no.~8, pp.~777--84, 2006.

\bibitem{RN490}
B.~Lu, ``Bdnf and activity-dependent synaptic modulation,'' {\em Learn Mem},
  vol.~10, no.~2, pp.~86--98, 2003.

\bibitem{RN494}
H.~Thoenen, ``Neurotrophins and activity-dependent plasticity,'' {\em Prog
  Brain Res}, vol.~128, pp.~183--91, 2000.

\bibitem{RN531}
L.~Pellerin, A.~K. Bouzier-Sore, A.~Aubert, S.~Serres, M.~Merle, R.~Costalat,
  and P.~J. Magistretti, ``Activity-dependent regulation of energy metabolism
  by astrocytes: an update,'' {\em Glia}, vol.~55, no.~12, pp.~1251--62, 2007.

\bibitem{RN533}
D.~T. Theodosis, D.~A. Poulain, and S.~H. Oliet, ``Activity-dependent
  structural and functional plasticity of astrocyte-neuron interactions,'' {\em
  Physiol Rev}, vol.~88, no.~3, pp.~983--1008, 2008.

\bibitem{RN606}
W.~S. Chung, L.~E. Clarke, G.~X. Wang, B.~K. Stafford, A.~Sher, C.~Chakraborty,
  J.~Joung, L.~C. Foo, A.~Thompson, C.~Chen, S.~J. Smith, and B.~A. Barres,
  ``Astrocytes mediate synapse elimination through megf10 and mertk pathways,''
  {\em Nature}, vol.~504, no.~7480, pp.~394--400, 2013.

\bibitem{RN488}
J.~L. Frost and D.~P. Schafer, ``Microglia: Architects of the developing
  nervous system,'' {\em Trends Cell Biol}, vol.~26, no.~8, pp.~587--597, 2016.

\bibitem{RN604}
D.~P. Schafer, E.~K. Lehrman, A.~G. Kautzman, R.~Koyama, A.~R. Mardinly,
  R.~Yamasaki, R.~M. Ransohoff, M.~E. Greenberg, B.~A. Barres, and B.~Stevens,
  ``Microglia sculpt postnatal neural circuits in an activity and
  complement-dependent manner,'' {\em Neuron}, vol.~74, no.~4, pp.~691--705,
  2012.

\bibitem{RN524}
M.~H. Canu, M.~Carnaud, F.~Picquet, and L.~Goutebroze, ``Activity-dependent
  regulation of myelin maintenance in the adult rat,'' {\em Brain Res},
  vol.~1252, pp.~45--51, 2009.

\bibitem{RN518}
R.~D. Fields, ``A new mechanism of nervous system plasticity:
  activity-dependent myelination,'' {\em Nat Rev Neurosci}, vol.~16, no.~12,
  pp.~756--67, 2015.

\bibitem{RN517}
M.~Monje, ``Myelin plasticity and nervous system function,'' {\em Annu Rev
  Neurosci}, vol.~41, pp.~61--76, 2018.

\bibitem{RN603}
L.~M. Boulanger, G.~S. Huh, and C.~J. Shatz, ``Neuronal plasticity and cellular
  immunity: shared molecular mechanisms,'' {\em Current Opinion in
  Neurobiology}, vol.~11, no.~5, pp.~568--578, 2001.

\bibitem{RN608}
M.~X. Li, M.~Jia, L.~X. Yang, V.~Dunlap, and P.~G. Nelson, ``Pre- and
  postsynaptic mechanisms in hebbian activity-dependent synapse modification,''
  {\em J Neurobiol}, vol.~52, no.~3, pp.~241--50, 2002.

\bibitem{RN607}
J.~R. Wilkerson, N.-P. Tsai, M.~A. Maksimova, H.~Wu, N.~P. Cabalo, K.~W.
  Loerwald, J.~B. Dictenberg, J.~R. Gibson, and K.~M. Huber, ``A role for
  dendritic mglur5-mediated local translation of arc/arg3. 1 in mef2-dependent
  synapse elimination,'' {\em Cell reports}, vol.~7, no.~5, pp.~1589--1600,
  2014.

\bibitem{RN727}
P.~B. Baltes, ``On the incomplete architecture of human ontogeny: Selection,
  optimization, and compensation as foundation of developmental theory,'' {\em
  American Psychologist}, vol.~52, no.~4, pp.~366--380, 1997.

\bibitem{RN162}
G.~M. Edelman, ``Neural darwinism: selection and reentrant signaling in higher
  brain function,'' {\em Neuron}, vol.~10, no.~2, pp.~115--25, 1993.

\bibitem{RN565}
G.~M. Edelman and G.~Tononi, {\em A universe of consciousness: How matter
  becomes imagination}.
\newblock Basic books, 2008.

\bibitem{RN356}
U.~L.~M. Lövdén, ``Brain plasticity in human lifespan development: The
  exploration–selection– refinement model,'' {\em Annual Review
  ofDevelopmental Psychology}, 2019.

\bibitem{RN1643}
J.~E. Niven and S.~B. Laughlin, ``Energy limitation as a selective pressure on
  the evolution of sensory systems,'' {\em J Exp Biol}, vol.~211, no.~Pt 11,
  pp.~1792--804, 2008.

\bibitem{RN573}
T.~Lomo, ``The discovery of long-term potentiation,'' {\em Philos Trans R Soc
  Lond B Biol Sci}, vol.~358, no.~1432, pp.~617--20, 2003.

\bibitem{RN433}
J.~C. Magee and C.~Grienberger, ``Synaptic plasticity forms and functions,''
  {\em Annu Rev Neurosci}, vol.~43, pp.~95--117, 2020.

\bibitem{RN576}
B.~E. Herring and R.~A. Nicoll, ``Long-term potentiation: From camkii to ampa
  receptor trafficking,'' {\em Annu Rev Physiol}, vol.~78, pp.~351--65, 2016.

\bibitem{RN575}
J.~D. Pita-Almenar, M.~S. Collado, C.~M. Colbert, and A.~Eskin, ``Different
  mechanisms exist for the plasticity of glutamate reuptake during early
  long-term potentiation (ltp) and late ltp,'' {\em J Neurosci}, vol.~26,
  no.~41, pp.~10461--71, 2006.

\bibitem{RN693}
R.~Druga, ``Neocortical inhibitory system,'' {\em Folia Biol (Praha)}, vol.~55,
  no.~6, pp.~201--17, 2009.

\bibitem{RN534}
Z.~J. Huang and A.~Paul, ``The diversity of gabaergic neurons and neural
  communication elements,'' {\em Nat Rev Neurosci}, vol.~20, no.~9,
  pp.~563--572, 2019.

\bibitem{RN364}
H.~Markram, M.~Toledo-Rodriguez, Y.~Wang, A.~Gupta, G.~Silberberg, and C.~Wu,
  ``Interneurons of the neocortical inhibitory system,'' {\em Nat Rev
  Neurosci}, vol.~5, no.~10, pp.~793--807, 2004.

\bibitem{RN406}
R.~Tremblay, S.~Lee, and B.~Rudy, ``Gabaergic interneurons in the neocortex:
  From cellular properties to circuits,'' {\em Neuron}, vol.~91, no.~2,
  pp.~260--92, 2016.

\bibitem{RN667}
M.~Okun and I.~Lampl, ``Instantaneous correlation of excitation and inhibition
  during ongoing and sensory-evoked activities,'' {\em Nat Neurosci}, vol.~11,
  no.~5, pp.~535--7, 2008.

\bibitem{RN622}
S.~M. Baca, A.~Marin-Burgin, D.~A. Wagenaar, and J.~Kristan, W.~B.,
  ``Widespread inhibition proportional to excitation controls the gain of a
  leech behavioral circuit,'' {\em Neuron}, vol.~57, no.~2, pp.~276--289, 2008.

\bibitem{RN621}
R.~Coultrip, R.~Granger, and G.~Lynch, ``A cortical model of winner-take-all
  competition via lateral inhibition,'' {\em Neural Networks}, vol.~5, no.~1,
  pp.~47--54, 1992.

\bibitem{RN616}
H.~K. Hartline and F.~Ratliff, ``Inhibitory interaction of receptor units in
  the eye of limulus,'' {\em J Gen Physiol}, vol.~40, no.~3, pp.~357--76, 1957.

\bibitem{RN619}
Y.~Yu, M.~Migliore, M.~L. Hines, and G.~M. Shepherd, ``Sparse coding and
  lateral inhibition arising from balanced and unbalanced dendrodendritic
  excitation and inhibition,'' {\em J Neurosci}, vol.~34, no.~41,
  pp.~13701--13, 2014.

\bibitem{RN242}
Z.~J. Huang, A.~Kirkwood, T.~Pizzorusso, V.~Porciatti, B.~Morales, M.~F. Bear,
  L.~Maffei, and S.~Tonegawa, ``Bdnf regulates the maturation of inhibition and
  the critical period of plasticity in mouse visual cortex,'' {\em Cell},
  vol.~98, no.~6, pp.~739--755, 1999.

\bibitem{RN620}
A.~C. Arevian, V.~Kapoor, and N.~N. Urban, ``Activity-dependent gating of
  lateral inhibition in the mouse olfactory bulb,'' {\em Nat Neurosci},
  vol.~11, no.~1, pp.~80--7, 2008.

\bibitem{RN639}
L.~F. Abbott, J.~A. Varela, K.~Sen, and S.~B. Nelson, ``Synaptic depression and
  cortical gain control,'' {\em Science}, vol.~275, no.~5297, pp.~220--4, 1997.

\bibitem{RN611}
J.~A. Varela, S.~Song, G.~G. Turrigiano, and S.~B. Nelson, ``Differential
  depression at excitatory and inhibitory synapses in visual cortex,'' {\em The
  Journal of Neuroscience}, vol.~19, no.~11, pp.~4293--4304, 1999.

\bibitem{RN712}
C.~Kapfer, A.~H. Seidl, H.~Schweizer, and B.~Grothe, ``Experience-dependent
  refinement of inhibitory inputs to auditory coincidence-detector neurons,''
  {\em Nat Neurosci}, vol.~5, no.~3, pp.~247--53, 2002.

\bibitem{RN159}
M.~Xue, B.~V. Atallah, and M.~Scanziani, ``Equalizing excitation-inhibition
  ratios across visual cortical neurons,'' {\em Nature}, vol.~511, no.~7511,
  pp.~596--600, 2014.

\bibitem{RN492}
N.~M. Bannon, M.~Chistiakova, and M.~Volgushev, ``Synaptic plasticity in
  cortical inhibitory neurons: What mechanisms may help to balance synaptic
  weight changes?,'' {\em Frontiers in Cellular Neuroscience}, vol.~14, 2020.

\bibitem{RN813}
R.~Azouz and C.~M. Gray, ``Dynamic spike threshold reveals a mechanism for
  synaptic coincidence detection in cortical neurons in vivo,'' {\em Proc Natl
  Acad Sci U S A}, vol.~97, no.~14, pp.~8110--5, 2000.

\bibitem{RN1639}
G.~Buzsaki, ``Neural syntax: cell assemblies, synapsembles, and readers,'' {\em
  Neuron}, vol.~68, no.~3, pp.~362--85, 2010.

\bibitem{RN589}
J.~A. Cardin, L.~A. Palmer, and D.~Contreras, ``Stimulus feature selectivity in
  excitatory and inhibitory neurons in primary visual cortex,'' {\em J
  Neurosci}, vol.~27, no.~39, pp.~10333--44, 2007.

\bibitem{RN601}
P.~B. Cook and J.~S. McReynolds, ``Lateral inhibition in the inner retina is
  important for spatial tuning of ganglion cells,'' {\em Nat Neurosci}, vol.~1,
  no.~8, pp.~714--9, 1998.

\bibitem{RN649}
S.~H. Lee, A.~C. Kwan, S.~Zhang, V.~Phoumthipphavong, J.~G. Flannery, S.~C.
  Masmanidis, H.~Taniguchi, Z.~J. Huang, F.~Zhang, E.~S. Boyden, K.~Deisseroth,
  and Y.~Dan, ``Activation of specific interneurons improves v1 feature
  selectivity and visual perception,'' {\em Nature}, vol.~488, no.~7411,
  pp.~379--83, 2012.

\bibitem{RN449}
S.~A. Josselyn and P.~W. Frankland, ``Memory allocation: Mechanisms and
  function,'' {\em Annu Rev Neurosci}, vol.~41, pp.~389--413, 2018.

\bibitem{RN447}
P.~Rao-Ruiz, J.~Yu, S.~A. Kushner, and S.~A. Josselyn, ``Neuronal competition:
  microcircuit mechanisms define the sparsity of the engram,'' {\em Curr Opin
  Neurobiol}, vol.~54, pp.~163--170, 2019.

\bibitem{RN450}
H.~K. Titley, N.~Brunel, and C.~Hansel, ``Toward a neurocentric view of
  learning,'' {\em Neuron}, vol.~95, no.~1, pp.~19--32, 2017.

\bibitem{RN697}
G.~M. Edelman and J.~A. Gally, ``Reentry: a key mechanism for integration of
  brain function,'' {\em Frontiers in integrative neuroscience}, vol.~7, p.~63,
  2013.

\bibitem{RN745}
V.~A. Lamme, H.~Super, and H.~Spekreijse, ``Feedforward, horizontal, and
  feedback processing in the visual cortex,'' {\em Curr Opin Neurobiol},
  vol.~8, no.~4, pp.~529--35, 1998.

\bibitem{lamme2000distinct}
V.~A. Lamme and P.~R. Roelfsema, ``The distinct modes of vision offered by
  feedforward and recurrent processing,'' {\em Trends in neurosciences},
  vol.~23, no.~11, pp.~571--579, 2000.

\bibitem{RN633}
P.~Girard, J.~Hupé, and J.~Bullier, ``Feedforward and feedback connections
  between areas v1 and v2 of the monkey have similar rapid conduction
  velocities,'' {\em Journal of Neurophysiology}, vol.~85, no.~3,
  pp.~1328--1331, 2001.

\bibitem{RN632}
J.~A. Hirsch and C.~D. Gilbert, ``Synaptic physiology of horizontal connections
  in the cat's visual cortex,'' {\em J Neurosci}, vol.~11, no.~6, pp.~1800--9,
  1991.

\bibitem{RN526}
A.~Messinger, L.~R. Squire, S.~M. Zola, and T.~D. Albright, ``Neuronal
  representations of stimulus associations develop in the temporal lobe during
  learning,'' {\em Proc Natl Acad Sci U S A}, vol.~98, no.~21, pp.~12239--44,
  2001.

\bibitem{RN1651}
M.~Tsodyks, T.~Kenet, A.~Grinvald, and A.~Arieli, ``Linking spontaneous
  activity of single cortical neurons and the underlying functional
  architecture,'' {\em Science}, vol.~286, no.~5446, pp.~1943--6, 1999.

\bibitem{RN628}
H.~Adesnik and M.~Scanziani, ``Lateral competition for cortical space by
  layer-specific horizontal circuits,'' {\em Nature}, vol.~464, no.~7292,
  pp.~1155--60, 2010.

\bibitem{RN1655}
G.~Buzsaki, {\em Rhythms of the Brain}.
\newblock Oxford university press, 2006.

\bibitem{RN635}
Y.~Dan and M.~M. Poo, ``Spike timing-dependent plasticity of neural circuits,''
  {\em Neuron}, vol.~44, no.~1, pp.~23--30, 2004.

\bibitem{RN634}
V.~Pawlak, J.~R. Wickens, A.~Kirkwood, and J.~N. Kerr, ``Timing is not
  everything: Neuromodulation opens the stdp gate,'' {\em Front Synaptic
  Neurosci}, vol.~2, p.~146, 2010.

\bibitem{RN768}
E.~T. Rolls, A.~Treves, and M.~J. Tovee, ``The representational capacity of the
  distributed encoding of information provided by populations of neurons in
  primate temporal visual cortex,'' {\em Exp Brain Res}, vol.~114, no.~1,
  pp.~149--62, 1997.

\bibitem{RN405}
Y.~Li, J.~B. Aimone, X.~Xu, E.~M. Callaway, and F.~H. Gage, ``Development of
  gabaergic inputs controls the contribution of maturing neurons to the adult
  hippocampal network,'' {\em Proc Natl Acad Sci U S A}, vol.~109, no.~11,
  pp.~4290--5, 2012.

\bibitem{RN843}
R.~Quian~Quiroga, ``Plugging in to human memory: Advantages, challenges, and
  insights from human single-neuron recordings,'' {\em Cell}, vol.~179, no.~5,
  pp.~1015--1032, 2019.

\bibitem{RN624}
D.~Purves, L.~E. White, and D.~R. Riddle, ``Is neural development darwinian?,''
  {\em Trends in neurosciences}, vol.~19, no.~11, pp.~460--464, 1996.

\bibitem{RN721}
A.~Gdalyahu, E.~Tring, P.~O. Polack, R.~Gruver, P.~Golshani, M.~S. Fanselow,
  A.~J. Silva, and J.~T. Trachtenberg, ``Associative fear learning enhances
  sparse network coding in primary sensory cortex,'' {\em Neuron}, vol.~75,
  no.~1, pp.~121--32, 2012.

\bibitem{RN734}
Y.~Yotsumoto, T.~Watanabe, and Y.~Sasaki, ``Different dynamics of performance
  and brain activation in the time course of perceptual learning,'' {\em
  Neuron}, vol.~57, no.~6, pp.~827--33, 2008.

\bibitem{RN733}
E.~Wenger, C.~Brozzoli, U.~Lindenberger, and M.~Lovden, ``Expansion and
  renormalization of human brain structure during skill acquisition,'' {\em
  Trends Cogn Sci}, vol.~21, no.~12, pp.~930--939, 2017.

\bibitem{RN758}
E.~T. Rolls and A.~Treves, ``The neuronal encoding of information in the
  brain,'' {\em Prog Neurobiol}, vol.~95, no.~3, pp.~448--90, 2011.

\bibitem{RN625}
J.~M. Crook, Z.~F. Kisvarday, and U.~T. Eysel, ``Evidence for a contribution of
  lateral inhibition to orientation tuning and direction selectivity in cat
  visual cortex: reversible inactivation of functionally characterized sites
  combined with neuroanatomical tracing techniques,'' {\em Eur J Neurosci},
  vol.~10, no.~6, pp.~2056--75, 1998.

\bibitem{RN626}
N.~J. Priebe and D.~Ferster, ``Inhibition, spike threshold, and stimulus
  selectivity in primary visual cortex,'' {\em Neuron}, vol.~57, no.~4,
  pp.~482--97, 2008.

\bibitem{RN756}
E.~De~Falco, M.~J. Ison, I.~Fried, and R.~Quian~Quiroga, ``Long-term coding of
  personal and universal associations underlying the memory web in the human
  brain,'' {\em Nat Commun}, vol.~7, p.~13408, 2016.

\bibitem{RN845}
P.~Fldik, ``Sparse coding in the primate cortex,'' {\em M Arbib the Handbook of
  Brain Theory \& Neural Networks}, 2002.

\bibitem{RN696}
B.~A. Olshausen and D.~J. Field, ``Sparse coding of sensory inputs,'' {\em Curr
  Opin Neurobiol}, vol.~14, no.~4, pp.~481--7, 2004.

\bibitem{RN894}
G.~Yang and W.~B. Gan, ``Sleep contributes to dendritic spine formation and
  elimination in the developing mouse somatosensory cortex,'' {\em Dev
  Neurobiol}, vol.~72, no.~11, pp.~1391--8, 2012.

\bibitem{RN895}
W.~Li, L.~Ma, G.~Yang, and W.~B. Gan, ``Rem sleep selectively prunes and
  maintains new synapses in development and learning,'' {\em Nat Neurosci},
  vol.~20, no.~3, pp.~427--437, 2017.

\bibitem{RN638}
H.~Markram and M.~Tsodyks, ``Redistribution of synaptic efficacy between
  neocortical pyramidal neurons,'' {\em Nature}, vol.~382, no.~6594,
  pp.~807--810, 1996.

\bibitem{RN698}
A.~M. Thomson, J.~Deuchars, and D.~C. West, ``Large, deep layer pyramid-pyramid
  single axon epsps in slices of rat motor cortex display paired pulse and
  frequency-dependent depression, mediated presynaptically and
  self-facilitation, mediated postsynaptically,'' {\em J Neurophysiol},
  vol.~70, no.~6, pp.~2354--69, 1993.

\bibitem{RN1653}
Y.~Ziv, L.~D. Burns, E.~D. Cocker, E.~O. Hamel, K.~K. Ghosh, L.~J. Kitch,
  A.~El~Gamal, and M.~J. Schnitzer, ``Long-term dynamics of ca1 hippocampal
  place codes,'' {\em Nat Neurosci}, vol.~16, no.~3, pp.~264--6, 2013.

\bibitem{RN690}
D.~Bavelier, D.~M. Levi, R.~W. Li, Y.~Dan, and T.~K. Hensch, ``Removing brakes
  on adult brain plasticity: from molecular to behavioral interventions,'' {\em
  J Neurosci}, vol.~30, no.~45, pp.~14964--71, 2010.

\bibitem{RN360}
R.~C. Froemke, ``Plasticity of cortical excitatory-inhibitory balance,'' {\em
  Annu Rev Neurosci}, vol.~38, pp.~195--219, 2015.

\bibitem{RN615}
T.~K. Hensch, ``Critical period plasticity in local cortical circuits,'' {\em
  Nat Rev Neurosci}, vol.~6, no.~11, pp.~877--88, 2005.

\bibitem{RN689}
A.~E. Takesian and T.~K. Hensch, ``Balancing plasticity/stability across brain
  development,'' {\em Prog Brain Res}, vol.~207, pp.~3--34, 2013.

\bibitem{RN400}
Liu and Guosong, ``Local structural balance and functional interaction of
  excitatory and inhibitory synapses in hippocampal dendrites,'' {\em Nature
  Neuroscience}, vol.~7, no.~4, pp.~373--379, 2004.

\bibitem{RN699}
A.~J. Peters, H.~Liu, and T.~Komiyama, ``Learning in the rodent motor cortex,''
  {\em Annu Rev Neurosci}, vol.~40, pp.~77--97, 2017.

\bibitem{RN407}
D.~van Versendaal, R.~Rajendran, M.~H. Saiepour, J.~Klooster, L.~Smit-Rigter,
  J.~P. Sommeijer, C.~I. De~Zeeuw, S.~B. Hofer, J.~A. Heimel, and C.~N. Levelt,
  ``Elimination of inhibitory synapses is a major component of adult ocular
  dominance plasticity,'' {\em Neuron}, vol.~74, no.~2, pp.~374--83, 2012.

\bibitem{RN728}
F.~Conde, J.~S. Lund, and D.~A. Lewis, ``The hierarchical development of monkey
  visual cortical regions as revealed by the maturation of
  parvalbumin-immunoreactive neurons,'' {\em Developmental Brain Research},
  vol.~96, no.~1-2, pp.~261--276, 1996.

\bibitem{dragoi2003}
G.~Dragoi, K.~D. Harris, and G.~Buzs{\'a}ki, ``Place representation within
  hippocampal networks is modified by long-term potentiation,'' {\em Neuron},
  vol.~39, no.~5, pp.~843--853, 2003.

\bibitem{RN700}
R.~Dawkins and N.~Davis, {\em The selfish gene}.
\newblock Macat Library, 2017.

\bibitem{RN871}
R.~Brette, ``Is coding a relevant metaphor for the brain?,'' {\em Behav Brain
  Sci}, vol.~42, p.~e215, 2018.

\bibitem{RN887}
S.~Harnad, ``The symbol grounding problem,'' {\em Physica D: Nonlinear
  Phenomena}, vol.~42, no.~1-3, pp.~335--346, 1990.

\bibitem{RN809}
C.~Blakemore and D.~E. Mitchell, ``Environmental modification of the visual
  cortex and the neural basis of learning and memory,'' {\em Nature}, vol.~241,
  no.~5390, pp.~467--8, 1973.

\bibitem{RN662}
I.~Gauthier, P.~Skudlarski, J.~C. Gore, and A.~W. Anderson, ``Expertise for
  cars and birds recruits brain areas involved in face recognition,'' {\em Nat
  Neurosci}, vol.~3, no.~2, pp.~191--7, 2000.

\bibitem{RN829}
I.~V. Viskontas, R.~Q. Quiroga, and I.~Fried, ``Human medial temporal lobe
  neurons respond preferentially to personally relevant images,'' {\em Proc
  Natl Acad Sci U S A}, vol.~106, no.~50, pp.~21329--34, 2009.

\bibitem{RN701}
H.~B. Barlow, ``Possible principles underlying the transformation of sensory
  messages,'' {\em Sensory communication}, vol.~1, no.~01, 1961.

\bibitem{RN472}
D.~C. Plaut and J.~L. McClelland, ``Locating object knowledge in the brain:
  comment on bowers's (2009) attempt to revive the grandmother cell
  hypothesis,'' {\em Psychol Rev}, vol.~117, no.~1, pp.~284--8, 2010.

\bibitem{RN499}
R.~Quian~Quiroga and G.~Kreiman, ``Measuring sparseness in the brain: comment
  on bowers (2009),'' {\em Psychol Rev}, vol.~117, no.~1, pp.~291--7, 2010.

\bibitem{RN868}
T.~Poggio, ``The levels of understanding framework, revised,'' {\em
  Perception}, vol.~41, no.~9, pp.~1017--23, 2012.

\bibitem{RN592}
R.~C. deCharms and A.~Zador, ``Neural representation and the cortical code,''
  {\em Annu Rev Neurosci}, vol.~23, pp.~613--47, 2000.

\bibitem{RN815}
C.~Keysers, D.~K. Xiao, P.~Foldiak, and D.~I. Perrett, ``The speed of sight,''
  {\em J Cogn Neurosci}, vol.~13, no.~1, pp.~90--101, 2001.

\bibitem{RN818}
D.~L. Sheinberg and N.~K. Logothetis, ``Noticing familiar objects in real world
  scenes: the role of temporal cortical neurons in natural vision,'' {\em J
  Neurosci}, vol.~21, no.~4, pp.~1340--50, 2001.

\bibitem{RN1547}
D.~L. Sheinberg and N.~K. Logothetis, ``Noticing familiar objects in real world
  scenes: the role of temporal cortical neurons in natural vision,'' {\em J
  Neurosci}, vol.~21, no.~4, pp.~1340--50, 2001.

\bibitem{RN259}
Y.~Elgersma and A.~J. Silva, ``Molecular mechanisms of synaptic plasticity and
  memory,'' {\em Curr Opin Neurobiol}, vol.~9, no.~2, pp.~209--13, 1999.

\bibitem{RN731}
M.~Fu, X.~Yu, J.~Lu, and Y.~Zuo, ``Repetitive motor learning induces
  coordinated formation of clustered dendritic spines in vivo,'' {\em Nature},
  vol.~483, no.~7387, pp.~92--5, 2012.

\bibitem{RN366}
M.~M. Poo, M.~Pignatelli, T.~J. Ryan, S.~Tonegawa, T.~Bonhoeffer, K.~C. Martin,
  A.~Rudenko, L.~H. Tsai, R.~W. Tsien, G.~Fishell, C.~Mullins, J.~T. Goncalves,
  M.~Shtrahman, S.~T. Johnston, F.~H. Gage, Y.~Dan, J.~Long, G.~Buzsaki, and
  C.~Stevens, ``What is memory? the present state of the engram,'' {\em BMC
  Biol}, vol.~14, p.~40, 2016.

\bibitem{RN735}
G.~Yang, F.~Pan, and W.~B. Gan, ``Stably maintained dendritic spines are
  associated with lifelong memories,'' {\em Nature}, vol.~462, no.~7275,
  pp.~920--4, 2009.

\bibitem{RN767}
T.~J. Gawne and B.~J. Richmond, ``How independent are the messages carried by
  adjacent inferior temporal cortical neurons?,'' {\em The Journal of
  Neuroscience}, vol.~13, no.~7, pp.~2758--2771, 1993.

\bibitem{RN892}
P.~M. Knutsen and E.~Ahissar, ``Orthogonal coding of object location,'' {\em
  Trends Neurosci}, vol.~32, no.~2, pp.~101--9, 2009.

\bibitem{RN641}
M.~Riesenhuber and T.~Poggio, ``Hierarchical models of object recognition in
  cortex,'' {\em Nat Neurosci}, vol.~2, no.~11, pp.~1019--25, 1999.

\bibitem{RN738}
M.~S. Gazzaniga, {\em The cognitive neurosciences}.
\newblock MIT press, 2009.

\bibitem{RN799}
G.~D. Logan and M.~J.~C. Crump, {\em Hierarchical Control of Cognitive
  Processes}, pp.~1--27.
\newblock Psychology of Learning and Motivation, 2011.

\bibitem{RN590}
H.~Freyhof, H.~Gruber, and A.~Ziegler, ``Expertise and hierarchical knowledge
  representation in chess,'' {\em Psychological Research}, vol.~54, no.~1,
  pp.~32--37, 1992.

\bibitem{RN742}
J.~Delgado, ``Integrative activity of the brain,'' {\em Yale Journal of Biology
  \& Medicine}, vol.~40, no.~4, pp.~334--335, 1968.

\bibitem{RN706}
W.~A. Freiwald, A.~K. Kreiter, and W.~Singer, ``Synchronization and assembly
  formation in the visual cortex,'' {\em Prog Brain Res}, vol.~130,
  pp.~111--40, 2001.

\bibitem{RN766}
L.~F. Abbott, E.~T. Rolls, and M.~J. Tovee, ``Representational capacity of face
  coding in monkeys,'' {\em Cereb Cortex}, vol.~6, no.~3, pp.~498--505, 1996.

\bibitem{RN743}
A.~P. Georgopoulos, A.~B. Schwartz, and R.~E. Kettner, ``Neuronal population
  coding of movement direction,'' {\em Science}, vol.~233, no.~4771,
  pp.~1416--9, 1986.

\bibitem{RN708}
W.~Singer, ``Consciousness and the structure of neuronal representations,''
  {\em Philos Trans R Soc Lond B Biol Sci}, vol.~353, no.~1377, pp.~1829--40,
  1998.

\bibitem{RN537}
J.~S. Bowers, N.~D. Martin, and E.~M. Gale, ``Researchers keep rejecting
  grandmother cells after running the wrong experiments: The issue is how
  familiar stimuli are identified,'' {\em Bioessays}, vol.~41, no.~8,
  p.~e1800248, 2019.

\bibitem{RN486}
C.~L. Buckley, C.~S. Kim, S.~McGregor, and A.~K. Seth, ``The free energy
  principle for action and perception: A mathematical review,'' {\em Journal of
  Mathematical Psychology}, vol.~81, pp.~55--79, 2017.

\bibitem{RN747}
T.~J. Sejnowski, ``Open questions about computation in cerebral cortex,'' {\em
  MIT Press}, 1986.

\bibitem{RN853}
N.~K. Logothetis, J.~Pauls, and T.~Poggio, ``Shape representation in the
  inferior temporal cortex of monkeys,'' {\em Curr Biol}, vol.~5, no.~5,
  pp.~552--63, 1995.

\bibitem{RN529}
E.~T. Rolls, G.~C. Baylis, M.~E. Hasselmo, and V.~Nalwa, ``The effect of
  learning on the face selective responses of neurons in the cortex in the
  superior temporal sulcus of the monkey,'' {\em Exp Brain Res}, vol.~76,
  no.~1, pp.~153--64, 1989.

\bibitem{RN654}
K.~J. Friston, R.~Rosch, T.~Parr, C.~Price, and H.~Bowman, ``Deep temporal
  models and active inference,'' {\em Neurosci Biobehav Rev}, vol.~90,
  pp.~486--501, 2018.

\bibitem{RN481}
R.~P.~N. Rao and D.~H. Ballard, ``Predictive coding in the visual cortex: a
  functional interpretation of some extra-classical receptive-field effects,''
  {\em Nature Neuroscience}, vol.~2, no.~1, pp.~79--87, 1999.

\bibitem{RN487}
K.~Friston, J.~Kilner, and L.~Harrison, ``A free energy principle for the
  brain,'' {\em J Physiol Paris}, vol.~100, no.~1-3, pp.~70--87, 2006.

\bibitem{RN749}
A.~Clark, ``Whatever next? predictive brains, situated agents, and the future
  of cognitive science,'' {\em Behav Brain Sci}, vol.~36, no.~3, pp.~181--204,
  2013.

\bibitem{RN418}
K.~Friston, ``The free-energy principle: a unified brain theory?,'' {\em Nat
  Rev Neurosci}, vol.~11, no.~2, pp.~127--38, 2010.

\bibitem{RN752}
M.~F. Panichello, O.~S. Cheung, and M.~Bar, ``Predictive feedback and conscious
  visual experience,'' {\em Front Psychol}, vol.~3, p.~620, 2012.

\bibitem{RN784}
A.~L. Roskies, ``The binding problem,'' {\em Neuron}, vol.~24, no.~1, pp.~7--9,
  1999.

\bibitem{RN560}
A.~Treisman, ``The binding problem,'' {\em Curr Opin Neurobiol}, vol.~6, no.~2,
  pp.~171--8, 1996.

\bibitem{RN764}
A.~Treisman, ``Preattentive processing in vision,'' {\em Computer Vision,
  Graphics, and Image Processing}, vol.~31, no.~2, pp.~156--177, 1985.

\bibitem{RN763}
A.~Treisman, ``Features and objects: the fourteenth bartlett memorial
  lecture,'' {\em Q J Exp Psychol A}, vol.~40, no.~2, pp.~201--37, 1988.

\bibitem{RN559}
A.~M. Treisman and G.~Gelade, ``A feature-integration theory of attention,''
  {\em Cognitive psychology}, vol.~12, no.~1, pp.~97--136, 1980.

\bibitem{RN508}
A.~K. Engel, P.~Fries, P.~Konig, M.~Brecht, and W.~Singer, ``Temporal binding,
  binocular rivalry, and consciousness,'' {\em Conscious Cogn}, vol.~8, no.~2,
  pp.~128--51, 1999.

\bibitem{RN420}
C.~v.~d. Malsburg, ``The correlation theory of brain function,'' {\em Models of
  Neural Networks}, 1994.

\bibitem{RN702}
W.~Singer and C.~M. Gray, ``Visual feature integration and the temporal
  correlation hypothesis,'' {\em Annu Rev Neurosci}, vol.~18, pp.~555--86,
  1995.

\bibitem{RN751}
A.~K. Seth, J.~L. McKinstry, G.~M. Edelman, and J.~L. Krichmar, ``Visual
  binding through reentrant connectivity and dynamic synchronization in a
  brain-based device,'' {\em Cereb Cortex}, vol.~14, no.~11, pp.~1185--99,
  2004.

\bibitem{RN506}
V.~Di~Lollo, ``The feature-binding problem is an ill-posed problem,'' {\em
  Trends in Cognitive Sciences}, vol.~16, no.~6, pp.~317--321, 2012.

\bibitem{RN789}
G.~M. Ghose and J.~Maunsell, ``Specialized representations in visual cortex: a
  role for binding?,'' {\em Neuron}, vol.~24, no.~1, pp.~79--85, 111--25, 1999.

\bibitem{RN507}
M.~N. Shadlen and J.~A. Movshon, ``Synchrony unbound: a critical evaluation of
  the temporal binding hypothesis,'' {\em Neuron}, vol.~24, no.~1, pp.~67--77,
  111--25, 1999.

\bibitem{RN793}
A.~Treisman, ``Feature binding, attention and object perception,'' {\em Philos
  Trans R Soc Lond B Biol Sci}, vol.~353, no.~1373, pp.~1295--306, 1998.

\bibitem{RN762}
J.~M. Wolfe and S.~C. Bennett, ``Preattentive object files: Shapeless bundles
  of basic features,'' {\em Vision research}, vol.~37, no.~1, pp.~25--43, 1997.

\bibitem{RN577}
F.~Baluch and L.~Itti, ``Mechanisms of top-down attention,'' {\em Trends
  Neurosci}, vol.~34, no.~4, pp.~210--24, 2011.

\bibitem{RN718}
G.~Deco and A.~Thiele, ``Attention: oscillations and neuropharmacology,'' {\em
  Eur J Neurosci}, vol.~30, no.~3, pp.~347--54, 2009.

\bibitem{RN547}
B.~Noudoost and T.~Moore, ``The role of neuromodulators in selective
  attention,'' {\em Trends Cogn Sci}, vol.~15, no.~12, pp.~585--91, 2011.

\bibitem{RN713}
J.~H. Reynolds and L.~Chelazzi, ``Attentional modulation of visual
  processing,'' {\em Annu Rev Neurosci}, vol.~27, pp.~611--47, 2004.

\bibitem{RN779}
H.~Zhou and R.~Desimone, ``Feature-based attention in the frontal eye field and
  area v4 during visual search,'' {\em Neuron}, vol.~70, no.~6, pp.~1205--17,
  2011.

\bibitem{RN792}
U.~Frey, Y.~Y. Huang, and E.~R. Kandel, ``Effects of camp simulate a late stage
  of ltp in hippocampal ca1 neurons,'' {\em Science}, vol.~260, no.~5114,
  pp.~1661--4, 1993.

\bibitem{RN795}
E.~P. Huang, ``Synaptic plasticity: Going through phases with ltp,'' {\em
  Current Biology}, vol.~8, no.~10, pp.~R350--R352, 1998.

\bibitem{RN705}
P.~Fries, S.~Neuenschwander, A.~K. Engel, R.~Goebel, and W.~Singer, ``Rapid
  feature selective neuronal synchronization through correlated latency
  shifting,'' {\em Nat Neurosci}, vol.~4, no.~2, pp.~194--200, 2001.

\bibitem{RN714}
P.~Fries, J.-H. Schröder, P.~R. Roelfsema, W.~Singer, and A.~K. Engel,
  ``Oscillatory neuronal synchronization in primary visual cortex as a
  correlate of stimulus selection,'' {\em The Journal of Neuroscience},
  vol.~22, no.~9, pp.~3739--3754, 2002.

\bibitem{RN1534}
P.~Fries, ``Neuronal gamma-band synchronization as a fundamental process in
  cortical computation,'' {\em Annu Rev Neurosci}, vol.~32, pp.~209--24, 2009.

\bibitem{RN797}
M.~Behrmann, M.~A. Peterson, M.~Moscovitch, and S.~Suzuki, ``Independent
  representation of parts and the relations between them: Evidence from
  integrative agnosia,'' {\em Journal of Experimental Psychology: Human
  Perception and Performance}, vol.~32, no.~5, pp.~1169--1184, 2006.

\bibitem{RN692}
P.~Fries, J.~H. Reynolds, A.~E. Rorie, and R.~Desimone, ``Modulation of
  oscillatory neuronal synchronization by selective visual attention,'' {\em
  Science}, vol.~291, no.~5508, pp.~1560--3, 2001.

\bibitem{RN716}
E.~I. Knudsen, ``Fundamental components of attention,'' {\em Annu Rev
  Neurosci}, vol.~30, pp.~57--78, 2007.

\bibitem{RN759}
K.~M. Shafritz, J.~C. Gore, and R.~Marois, ``The role of the parietal cortex in
  visual feature binding,'' {\em Proc Natl Acad Sci U S A}, vol.~99, no.~16,
  pp.~10917--22, 2002.

\bibitem{RN805}
M.~M. Chun, J.~D. Golomb, and N.~B. Turk-Browne, ``A taxonomy of external and
  internal attention,'' {\em Annu Rev Psychol}, vol.~62, pp.~73--101, 2011.

\bibitem{RN782}
C.~M. Gray, ``The temporal correlation hypothesis of visual feature
  integration: still alive and well,'' {\em Neuron}, vol.~24, no.~1,
  pp.~31--47, 111--25, 1999.

\bibitem{RN293}
V.~Colizza, A.~Flammini, M.~A. Serrano, and A.~Vespignani, ``Detecting
  rich-club ordering in complex networks,'' {\em Nature Physics}, vol.~2,
  no.~2, pp.~110--115, 2006.

\bibitem{RN322}
N.~A. Crossley, A.~Mechelli, P.~E. Vertes, T.~T. Winton-Brown, A.~X. Patel,
  C.~E. Ginestet, P.~McGuire, and E.~T. Bullmore, ``Cognitive relevance of the
  community structure of the human brain functional coactivation network,''
  {\em Proc Natl Acad Sci U S A}, vol.~110, no.~28, pp.~11583--8, 2013.

\bibitem{RN326}
G.~Zamora-Lopez, C.~Zhou, and J.~Kurths, ``Cortical hubs form a module for
  multisensory integration on top of the hierarchy of cortical networks,'' {\em
  Front Neuroinform}, vol.~4, p.~1, 2010.

\bibitem{RN284}
E.~Bullmore and O.~Sporns, ``The economy of brain network organization,'' {\em
  Nat Rev Neurosci}, vol.~13, no.~5, pp.~336--49, 2012.

\bibitem{RN338}
G.~Collin, O.~Sporns, R.~C. Mandl, and M.~P. van~den Heuvel, ``Structural and
  functional aspects relating to cost and benefit of rich club organization in
  the human cerebral cortex,'' {\em Cereb Cortex}, vol.~24, no.~9,
  pp.~2258--67, 2014.

\bibitem{RN801}
G.~D. Logan, ``Toward an instance theory of automatization,'' {\em
  Psychological Review}, vol.~95, no.~4, pp.~492--527, 1988.

\bibitem{RN806}
A.~Treisman, A.~Vieira, and A.~Hayes, ``Automaticity and preattentive
  processing,'' {\em The American Journal of Psychology}, vol.~105, no.~2,
  1992.

\bibitem{RN270}
D.~Hassabis, D.~Kumaran, C.~Summerfield, and M.~Botvinick,
  ``Neuroscience-inspired artificial intelligence,'' {\em Neuron}, vol.~95,
  no.~2, pp.~245--258, 2017.

\bibitem{RN462}
S.~Pinker and A.~Prince, ``On language and connectionism: analysis of a
  parallel distributed processing model of language acquisition,'' {\em
  Cognition}, vol.~28, no.~1-2, pp.~73--193, 1988.

\bibitem{RN595}
G.~Buzsaki, ``The brain-cognitive behavior problem: A retrospective,'' {\em
  eNeuro}, vol.~7, no.~4, 2020.

\bibitem{RN737}
P.~S. Churchland and T.~J. Sejnowski, ``Perspectives on cognitive
  neuroscience,'' {\em Science}, vol.~242, no.~4879, pp.~741--5, 1988.

\bibitem{RN570}
K.~Friston, ``Beyond phrenology: what can neuroimaging tell us about
  distributed circuitry?,'' {\em Annu Rev Neurosci}, vol.~25, pp.~221--50,
  2002.

\bibitem{RN594}
D.~Poeppel and F.~Adolfi, ``Against the epistemological primacy of the
  hardware: The brain from inside out, turned upside down,'' {\em eNeuro},
  vol.~7, no.~4, 2020.

\bibitem{RN862}
W.~T. Fitch, ``Toward a computational framework for cognitive biology: unifying
  approaches from cognitive neuroscience and comparative cognition,'' {\em Phys
  Life Rev}, vol.~11, no.~3, pp.~329--64, 2014.

\bibitem{fitch2008nano}
W.~T. Fitch, ``Nano-intentionality: a defense of intrinsic intentionality,''
  {\em Biology \& Philosophy}, vol.~23, no.~2, pp.~157--177, 2008.

\bibitem{RN826}
E.~O. Wilson, {\em Consilience: The unity of knowledge}, vol.~31.
\newblock Vintage, 1999.

\end{thebibliography}

\newpage

\end{document}